\documentclass[12pt,twoside,a4paper]{article}
\usepackage{graphicx}
\usepackage{amsmath}
\usepackage{url}
\usepackage{fancyvrb}
\usepackage[noend]{algorithmic}
\usepackage{algorithm}
\usepackage{multirow}
\usepackage{listing}
\usepackage{epsfig,endnotes}
\usepackage{amsmath}

\usepackage{times,amsmath,epsfig}
\usepackage[T1]{fontenc}
\usepackage[utf8]{inputenc}
\usepackage{listings}
\usepackage{listingsutf8}
\usepackage{parskip}
\usepackage{float}
\usepackage{caption}
\usepackage{textcomp}
\usepackage{graphicx}
\usepackage{epsfig}
\usepackage{amssymb}
\usepackage{longtable}
\usepackage{combelow}

 \floatstyle{ruled}
\newfloat{algorithm}{tbp}{loa}
\floatname{algorithm}{Algorithm}

\usepackage{url}
\usepackage{makeidx}

\makeatletter
\def\ext@lstlisting{lol}
\def\ext@figure{lol}
\AtBeginDocument{%
 \let\l@lstlisting\l@figure
  \let\c@lstlisting\c@figure

}
\makeatother

\usepackage{xcolor}

\colorlet{punct}{red!60!black}
\definecolor{background}{HTML}{EEEEEE}
\definecolor{delim}{RGB}{20,105,176}
\colorlet{numb}{magenta!60!black}

\lstdefinelanguage{json}{
    basicstyle=\normalfont\ttfamily,
    stepnumber=1,
    numbersep=8pt,
    showstringspaces=false,
    breaklines=true,
    frame=lines,
    backgroundcolor=\color{background},
    literate=
     *{0}{{{\color{numb}0}}}{1}
      {1}{{{\color{numb}1}}}{1}
      {2}{{{\color{numb}2}}}{1}
      {3}{{{\color{numb}3}}}{1}
      {4}{{{\color{numb}4}}}{1}
      {5}{{{\color{numb}5}}}{1}
      {6}{{{\color{numb}6}}}{1}
      {7}{{{\color{numb}7}}}{1}
      {8}{{{\color{numb}8}}}{1}
      {9}{{{\color{numb}9}}}{1}
      {:}{{{\color{punct}{:}}}}{1}
      {,}{{{\color{punct}{,}}}}{1}
      {\{}{{{\color{delim}{\{}}}}{1}
      {\}}{{{\color{delim}{\}}}}}{1}
      {[}{{{\color{delim}{[}}}}{1}
      {]}{{{\color{delim}{]}}}}{1},
}

\usepackage{color}
\definecolor{gray}{rgb}{0.4,0.4,0.4}
\definecolor{darkblue}{rgb}{0.0,0.0,0.6}
\definecolor{cyan}{rgb}{0.0,0.6,0.6}

\lstdefinelanguage{xml}
{
  morestring=[b]",
  morestring=[s]{>}{<},
  morecomment=[s]{<?}{?>},
  stringstyle=\color{black},
  identifierstyle=\color{darkblue},
  keywordstyle=\color{cyan},
  morekeywords={xmlns,version,type}
}

\makeindex

\begin{document}


\date{}

\title{\Large \bf Web Data Knowledge Extraction}

\author{
{\rm Juan M. Tirado, Ovidiu \cb{S}erban, Qiang Guo, and Eiko Yoneki}\\
University of Cambridge Computer Laboratory\\
{\small{\{jmt78,ovidiu.serban,qiang.guo,eiko.yoneki\}@cl.cam.ac.uk}}\\
Cambridge, United Kingdom\\
} 

\maketitle
\begin{abstract}
A constantly growing amount of information is available through the web. Unfortunately, extracting useful content from this massive amount of data still remains an open issue. The lack of standard data models and structures forces developers to create ad-hoc solutions from the scratch. The figure of the expert is still needed in many situations where developers do not have the correct background knowledge. This forces developers to spend time acquiring the needed background from the expert. In other directions, there are promising solutions employing machine learning techniques. However, increasing accuracy requires an increase in system complexity that cannot be endured in many projects.

In this work, we approach the web knowledge extraction problem using an expert-centric methodology. This methodology defines a set of configurable, extendible and independent components that permit the reuse of large pieces of code among projects. Our methodology differs from similar solutions in its expert-driven design. This design, makes it possible for subject-matter expert to drive the knowledge extraction for a given set of documents. Additionally, we propose the utilization of machine assisted solutions that guide the expert during this process. To demonstrate the capabilities of our methodology, we present a real use case scenario in which public procurement data is extracted from the web-based repositories of several public institutions across Europe. We provide insightful details about the challenges we had to deal with in this use case and additional discussions about how to apply our methodology.

\end{abstract}

\newpage
\tableofcontents
\lstlistoflistings
\listoftables
\newpage
\maketitle

\section{Introduction}

The explosion of the Internet, the web and the social networks have created a massive repository of available information. Society is producing data at a pace never seen before, and this trend seems set to continue with the adoption of new paradigms such as the Internet of Things (IoT). Although existing technologies simplify access to a large amount of data, the heterogeneity of formats, the lack of APIs, the Deep Web and the idiosyncrasie of data sources can make very difficult the acquisition and processing of these data.

Knowledge extraction is a major task in many companies and research projects that demand data allocated in the web in order to store it, analyse it or simply sell it to third parties. This task requires an understanding of the data layout and what has to be extracted. In some cases, the utilization of metadata or data model descriptions may help to understand the structure of the data. Unfortunately, this information is not available in most cases.

Most knowledge extraction is done in ad-hoc solutions designed from scratch. Normally these solutions comprise the acquisition, parsing, transformation and storage of the data. The process is normally conducted by developers, with a certain programming background. The developers have to deal with two different problems: the technical complexity of parsing a given document and understanding the semantics of the information contained in that document. Unfortunately, certain areas of knowledge may require subject-matter to identify these semantics. Normally, the developers work in conjunction with the expert as they may not have any technical background. This forces the developers to spend precious time absorbing the knowledge of the expert.

Apart from the technical difficulties of the aforementioned elements, the most difficult task is the knowledge extraction from the expert. We can consider the expert the keystone of the whole process. In this work, we approach the web knowledge extraction problem using an expert-centric methodology. Following this idea, the whole knowledge extraction task should be designed around the expert and her knowledge. From data acquisition to knowledge extraction, the expert is assisted by a set of tools that help her through the process with minimal intervention from the developers. Our methodology has the following features:

\begin{enumerate}
	\item \textbf{Expert-centric design}. The expert is the main actor of the web knowledge extraction process and she can eventually drive a full extraction pipeline with a minimal IT background.
	\item \textbf{Machine-assisted}. In many cases, data can be significantly complex (optional fields, repetitive structures, hierarchies, etc.). We expect machine learning solutions to assist experts in the process, making it possible to enhance and simplify the whole task.
	\item \textbf{Reusable}. Many of the tasks and subtasks that comprise knowledge extraction are repetitive. The definition of a framework can make possible the definition of common and reusable routines.
	\item \textbf{Generic solution}. Developing an ad-hoc solution makes it  difficult to maintain over time. However, a black-box approach where the behaviour of the system depends on a set of inputs and outputs reduces the whole problem to a pipeline design. This improves the maintainability of the code and it makes it possible to focus the effort on improving/creating black boxes.
	\item \textbf{Configurable and extensible}. We are aware of the complexity of dealing with a full-featured solution for any existing data source. For this reason, we consider that any methodology may take into consideration the possibility of developing additional extensions that allow easy adaptation to new use cases.
	\item \textbf{Format independent}. Any data acquisition strategy must be independent of the format of the incoming data (HTML, XML, CSV, TXT, etc.)
	\item \textbf{Database independent}. The acquisition process must be independent of the database destination and therefore, transparent for the expert.
\end{enumerate}

We believe that our methodology sets a framework for open discussion and reframes the web knowledge extraction problem from a different perspective. Our work differs from previous solutions in the following points:
\begin{description}
	\item[i)] We present a methodology that addresses the entire web knowledge acquisition life cycle.
	\item[ii)] We present, discuss and implement each of the elements that form our methodology.
	\item[iii)] We explore the limits of our methodology in a real use case and discuss the difficulties to be considered in every stage of the project.
\end{description}

The current document is organized as follows. We introduce the web data knowledge extraction problem in Section~\ref{sec:background} followed by a summary of the state of the art regarding this work in Section~\ref{sec:relatedwork}. We overview our methodology in Section~\ref{sec:systemoverview} and describe its components in Sections~\ref{sec:webcrawling}, \ref{sec:htmlxml}, \ref{sec:mapping} and \ref{sec:xmlpipeline}. We describe our validation in Section~\ref{sec:validation}. We describe a use case using our methodology in Section~\ref{sec:htmlprocurement} and conclude in Section~\ref{sec:conclusion}.

\newpage

\section{Background}
\label{sec:background}

The web requires an interoperable language with a syntax that permits the presentation of data to the user while maintaining a certain layout. HTML has become the language of the web and therefore, the most adopted solution. However, HTML does not provide by default any mechanism that facilitates the automatic analysis of existing documents. This limitation makes impossible to differentiate the content from the layout and the data semantics.

Several standards such as RDF, RDFS and OWL were developed to provide a common syntax to define data models. These solutions permit the definition of ontologies that support queries. These technologies are not usually understood by developers who initially ignore the semantic annotation process during the design of HTML pages. In order to simplify this problem, a more recent approach called Schema~\cite{schema_org} defines a vocabulary of concepts such as people, places, events and products that permit the annotation of data contained in an HTML document. This makes possible to stablish a relationship between the content and any existing schema bringing semantics to the web. 

The design of web pages may hide the data from the existing web search engines. The utilization of dynamic content, CAPTCHAS, private webs, scripted pages or unlinked content among others gives rise to the Deep Web~\cite{Raghavan:2001}. A simple example is a utilization of web pages that performs search queries against a database. The information contained in the database cannot be indexed by a search engine, as this requires the engine to interact with the search form, set the search parameters and understand the semantics of the returned data. Commercial search engines such as Google, Bing or DuckDuckGo design their tools with a clear emphasis in indexing what is called the surface web. This lead us to think that most of the information contained in the web is non-indexed. This problem has been recently highlighted by the DARPA's MEMEX project~\cite{memex} which shows an attempt to index the information contained in the Deep Web.

Processing the data contained in any of these web pages implies a certain degree of human interaction (filling a search form, interacting with a script, etc.). Once the raw data is downloaded it is transformed into some sort of indexable format that can be stored into a database for later analysis. When tackling the knowledge extraction problem most solutions are designed from scratch, dealing with data extraction, parsing and the storage. In the case of multiple data sources the complexity of the problem grows until it becomes unfeasible. 

We can identify three main actors in any knowledge extraction problem. (i) A data source containing relevant information (e.g. a web page). (ii) A destination database designed to store the data (e.g. MySQL). (iii) An expert that can determine how to transform data in (i) into (ii). The transformation between (i) and (ii) can be considered to be done automatically. Independently of the level of automation, the role of the expert is required in order to insert some initial semantics about the data prior the extraction can start. Additionally, the expert is responsible for defining whether the data extraction is correct or not. A fourth actor (iv) could be the developers. We define developer as a professional that can contribute to the development of the solution in a technical sense. The developer is intended to write the code, routines and programmes that prepare the data to be processed.


When dealing with any web knowledge extraction project there are several aspects to be considered. Based on our experience (see Section~\ref{sec:htmlprocurement}) we can enumerate some aspects normally ignored in existing designs:

\paragraph{Multiple data sources:} Most of the approaches dealing with data extraction plan to deal with just a few data sources, similar in design (e.g. Amazon product pages, Wikipedia, etc.). Our case study does not have a central authority, with a uniform design, where all the documents are archived. The system has to take into account that data sources do not have a unique design, mostly inconsistent within the source, and multiple sources of information.

\paragraph{Multilingual approach:} Most of the existing solutions only consider data sources written in a single language. A multilingual set of data sources or even data sources using multiple languages simultaneously. E.g. when dealing with extraction from European government related institutions, all the languages of the member countries are considered official and can be used to exchange documents. In most cases, member and partner countries use their official state language to publish any EU related document and the official, central, reports are available in English, French or German, as most frequently used languages, by number of speakers. This situation is encountered in any global organization and state unions with more than one spoken language.

\paragraph{Data extraction from various file format:} HTML is the most common data format. Actually, the adoption of HTML5 seems to strengthen its popularity. However, other formats such as XML, DOC or PDF may be present. For HTML and XML, various parsers exists, which preserve the formatting structure as well, so it can be used afterwards. DOC and PDF parsers are more difficult to find, and most of them extract the text data, without any formatting. This makes automatic data extraction more difficult. In some extreme cases, the files are archived into various formats (e.g. ZIP, RAR, self-extracting RAR) and a file pre-processing component has to be designed to extract the data.

\paragraph{Structured against unstructured:} Structured data has a rigorous schema definition, where all the constituents are defined by data types and linked to other elements of the data structure. By this definition, XML and, CSV are considered semi-structured, if the schema of the data is present (header in case of CSV files). In the current web environment, most of the documents fit into the unstructured or semi-structured category. Moreover, some XML documents are available, but the schema is not properly defined or unavailable. In some cases we can even find non properly defined XML structures.

\paragraph{Data format problems:} When dealing with legacy systems, designed at moments in time when the data standards were not established, the data format becomes a very big issue. In the web environment, some of the semi-structured documents may not follow their own schema. Moreover, data encoding is also a problem, which depends on the underlying system. In the current web world, UTF-8 is a widely used standard, but in some cases, the servers send a different encoding from the actual encoding of the file.

\paragraph{Reusable repetitive operations and continuous updates:} The system design has to take into account that data that originated from the selected data sources evolves in time, meaning that more documents may be added, but also the design of the extraction process may be changed due to underlying structural changes of the document. Most of the operations executed into the extraction pipeline can be reused across domains and data sources, therefore, the pipeline need to be reusable and customisable. Moreover, these operations need to be executed in a continuous and repetitive manner without much human intervention.

\newpage

\section{Related Work}
\label{sec:relatedwork}

\subsection{Crawling}

The existence of the Deep Web was documented back in 1998~\cite{Lawrence:98}. Since then, some works have presented solutions to automatically access and index these data. Raghavan and Garcia-Molina~\cite{Raghavan:2001} design a set of modules that permit to populate databases with existing data stored behind a search form. Their approach permits to manually define a set of labels that can be used to identify a valuable search form. If a valuable search form is detected, then the system launches the query and populates a database after labelling relevant data. Zerfos et al.~\cite{Zerfos:05} try to solve the same problem with a particular emphasis in discovering the maximum number of documents in systems only accessible through a search form. Shestakov et al.~\cite{Shestakov:2005} design a domain specific language called DEQUE that permits to transform SQL style queries into HTML form queries. Google presented its own solution~\cite{Madhavan:2008} that aims at finding the most informative queries in order to reduce the traffic while augmenting the scalability of the crawler.

Other family of crawling solutions permit the user to define ad-hoc the crawling strategy. Scrapy~\cite{scrapy} is a Python library designed to simplify the crawling process. Other solutions such as the Nutch project~\cite{nutch} are more oriented to large scalable crawlers on top of the Hadoop infrastructure and permit distributed crawling strategies. Recently an increasing number of SaaS (Software As A Service) solutions offer online interfaces to define the crawling strategies~\cite{scrapinghub,mozenda}. These platforms normally integrate plugins into the browser interface that permit to identify what items from a web page should be crawled. The simplification of the crawling process might limit the definition of the crawling policies. However, the utilization of cloud resources makes possible the deployment of large crawling solutions in a small time.

\subsection{Knowledge extraction}

Knowledge extraction is the creation of knowledge from structured or unstructured data sources. For a given set of documents we want to build knowledge stored in some data warehouse (typically a database or an ontological structure). A straight forward solution is to design an ad-hoc parser that processes the documents and extract data used to populate some data model. This approach can be sufficient for small amounts of data using known data structures. 

The simplest solution, adopted by many projects is using XQuery \cite{xquery} or regular expressions to extract the exact path of the targeted element. This approach is not very robust to structural changes of the document template. Another popular approach is using EXtensible Stylesheet Language Transformations (XSLT) \cite{xslt}, which provides a unified syntax to write transformation rules between XML compatible languages. In the basic form, HTML is mainly XML compatible, therefore this approach can be applied to HTML. This approach is more robust than XQuery to structural changes, but it is usually very difficult to debug. Another simple technique, very practical in small projects is HTML tidy \cite{htmltidy} or HTML simplification. This is able to simplify the HTML syntax by removing all the elements apart from the basic HTML syntax and formatting, and has proven effective in clean tabular approaches.   

In other scenarios, only the utilization of more sophisticated solutions can be a reasonable choice.

The problem of knowledge extraction has been intensively explored in the recent years. A first family of solutions has explored the utilization of domain specific languages to define how data should be extracted. Solutions like those presented in~\cite{Shen:2007,Krishnamurthy:2008} use declarative information extraction language to define data extraction plans. Similarly~\cite{Nakashole:2011} uses sets of rules to define the extraction. In these solutions the quality of the extraction is particularly dependant on the skills of the operators to define the extraction rules. A second family of solutions explores the utilization of machine learning techniques in order to improve the information extraction. These solutions are based on the utilization of inference models that try to construct relationships for a given dataset~\cite{Dong:2014,Etzioni:2004,Carlson:2010}. 

Liu et al \cite{liu2004editorial} presents a survey of existing Web Content mining methods, with focus on automated and machine learning approaches. In their classification, the structured and unstructured data extraction approaches, are very similar as concepts to our current problem. 

Barathi \cite{barathi2014structured} and Arasu, Garcia-Molina \cite{arasu2003extracting} use an automated iterative process to build formal template description of a series of documents, using a single, uniform, template. This approach can automatically detect repetitive structures inside a single template and create an approximation of the page structure, while being able to extract any relevant data. IEPAD \cite{chang2001iepad} uses a similar approach and reduces the complexity of the problem by grouping the HTML element tags into various categories. The IEPAD results is a PAT tree, which approximates a single, uniform, document template. Wang \cite{wang2004information} extends this approach, by comparing the tree similarity obtained by building multiple HTML element trees.

One popular framework that unifies the machine learning and dictionary-based approaches is GATE \cite{cunningham2013getting}. This toolkit provides a full framework for annotation, building dictionaries for Named Entities and various Natural Language Processing and Machine Learning techniques, very useful to build supervised approaches in Data Mining.    

Recently, the DeepDive~\cite{Shin:2015} framework has attracted a lot of attention form the research community. DeepDive uses a set of predefined rules to infer relationships between entities identified in . The final database creation is an iterative supervised loop where the operator can guide the machine learning process identifying the errors committed by the system. In a similar way Google's Knowledge Vault~\cite{Dong:2014} builds relationships using RDF triplets. This approach shows some resemblances with DeepDive with a clear emphasis in data scalability.

\newpage

\section{Methodology overview}
\label{sec:systemoverview}

\begin{figure}[H]
	\centering
	\includegraphics[width=\linewidth]{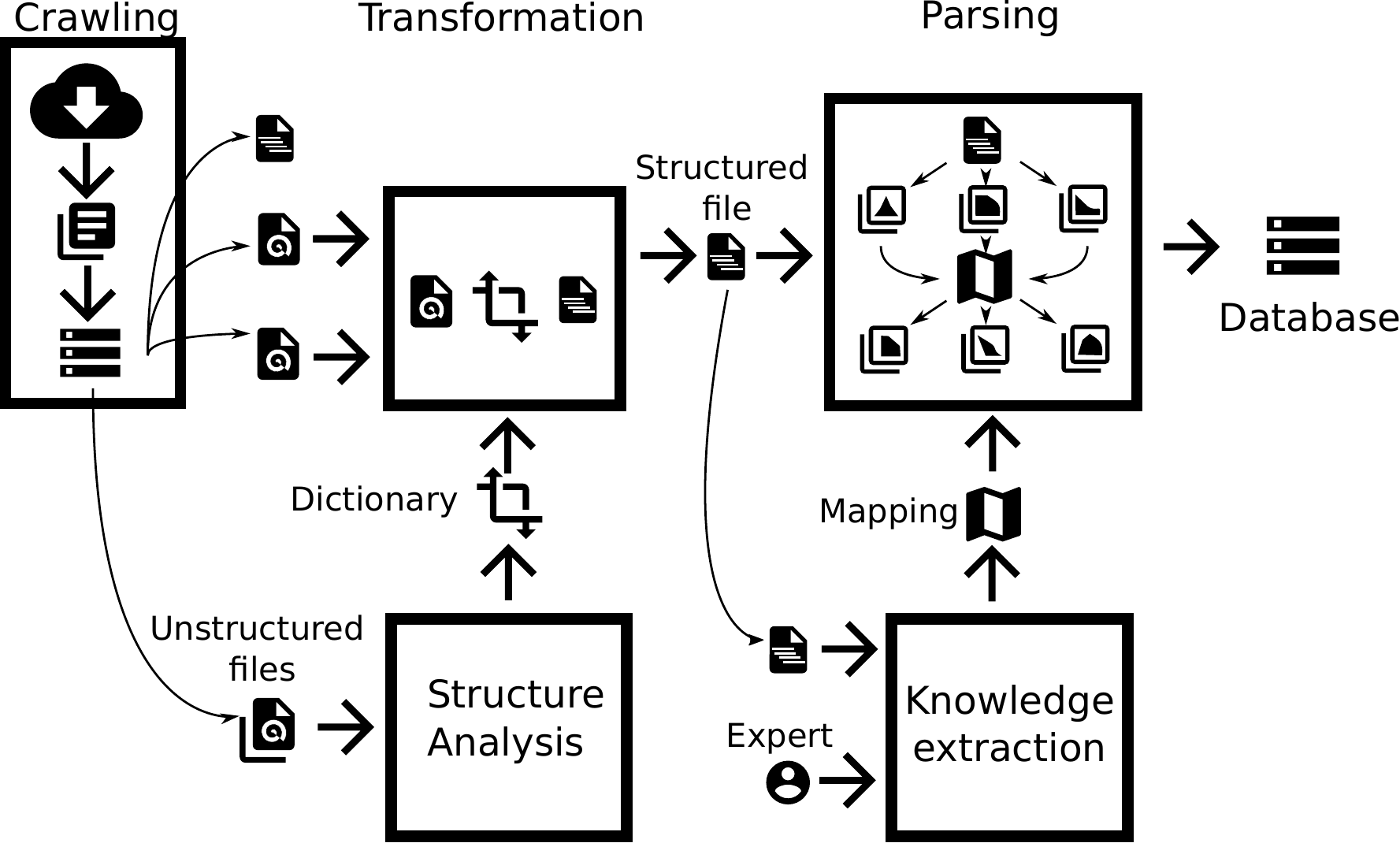} 
	\caption{Methodology overview}
	\label{fig:system_overview}
\end{figure}

We propose a machine-assisted expert-centric methodology to cope with the problem of web data knowledge extraction. Our methodology stresses the idea of the expert as the keystone of an iterative process that transforms existing raw data into a knowledge database. Figure~\ref{fig:system_overview} describes the different components configuring our methodology. 

\begin{itemize}
	\item \textbf{Crawling}. The crawling component is designed to collect the data from the original data sources for its later processing. The crawled elements are stored for its later processing into a secondary storage system such as a database or a file system.
	
	\item \textbf{Transformation}. The crawled elements may appear in different data formats (HTML, XML, json, txt, CSV, etc.) and in many cases the file to be processed can be found in a compressed format. In some cases, such as XML, the data structure is self-descriptive simplifying the parsing process, and therefore the data extraction. The transformation component converts raw data extracted from the original data source into an XML file (structured file in Figure~\ref{fig:system_overview}) containing the same information. In order to perform this transformation, this component employs a dictionary input which is created by the structure analysis component.
	
	\item \textbf{Structure analysis}. File formats such as HTML, CSV or TXT may have a structured content even if the original type is not self-descriptive. The structure analysis component permits to identify the structure of the elements contained into these files by analysing dataset samples. The final result is a dictionary containing a description about the elements that have to be considered as pieces of information during the transformation process. This module is language and format independent which permits its reuse without ad-hoc coding.
	
	\item \textbf{Knowledge extraction}. The final destination of the crawled information is a database. The schema of this database may differ from the original crawled data and requires a logical transformation. Because the crawled elements could have been obtained from different data sources, in different languages and formats it is necessary human intervention to drive the conversion between the original raw data and the final database destination. This knowledge extraction process, is a guided process where a human expert uses structured files (result of the transformation component) to define a mapping between the original data and the destination database. 
	
	\item \textbf{Parsing}. The parsing component uses the expert defined mapping to populate the destination database. It maps the semantics in the original data source to the common format represented in the target database. This component is completely reusable and only depends on the input data, and the mapping defined by the expert in order to carry out the data transformation. 
	
\end{itemize}

\subsection{Expert-centric operational approach}

\begin{figure}[H]
	\centering
	\includegraphics[width=0.6\linewidth]{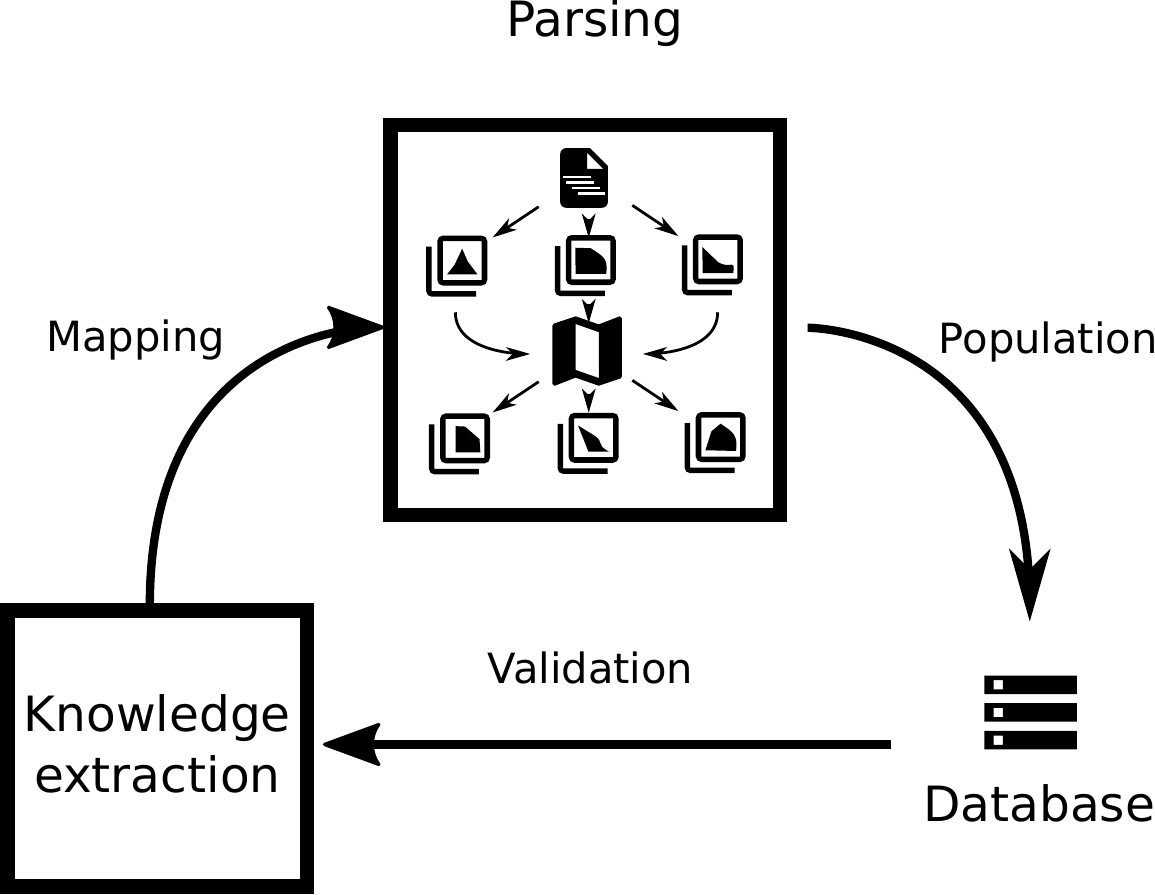} 
	\caption{Expert-centric operation cycle}
	\label{fig:system_iteration}
\end{figure}

Our methodology works in an expert-centric manner. The expert can operate a complete pipeline that includes mapping, populating a database and validating the parsed content. This operation is iteratively repeated (Figure~\ref{fig:system_iteration}) until the expert is satisfied with the results. Assuming an existing crawled collection of documents, the expert performs an initial mapping using the knowledge extraction component. This initial mapping is used by the parsing component and populates the target database. A validation of the contents into the database shall permit the expert to identify potential improvements in her mapping. By iteratively correcting the mapping and parsing/populating the database, the expert (or other operator) may decide when the quality of the target database is good enough. A more detailed discussion about validation is given in Section~\ref{sec:validation}.

The user can be assisted by machine-based solutions that facilitate work of the expert. The iterative nature of our solution may sound repetitive and even unfeasible. However, our experience shows that experts with a deep understanding of the data work iteratively in order to define the data mapping. The iterative approach follows a natural procedure of trial and error with many repetitive tasks such as finding keys describing a certain value and their equivalent into the destination database. Our system provides the expert with abstraction tools that hide implementation details to the experts and minimize the interaction with developers. In an ideal use case scenario the expert could define a complete pipeline without the intervention of the developers.

\newpage

\section{Web crawling}
\label{sec:webcrawling}

The web crawling is the first step into the data acquisition process. The elements to be crawled may vary depending on the data source. In a simplified scenario, we can assume that we aim at crawling web pages. How to crawl these web pages depends on the design of the data source. In some cases the pages are easily accessible through a single URL or can be obtained after fulfilling a search form. We enumerate possible scenarios to deal with:

\begin{itemize}
\item \textbf{Identification by public ID}. In this scenario every element to be crawled is uniquely identified by a URL that contains a unique identifier. If the generation of the identifiers is known it is possible to statically generate a list of possible URLs to be requested.
\item \textbf{Identification by unknown ID}. Similar to the previous scenario. However, how the identifiers are generated is unknown. In this case, the identifiers have to be initially extracted from the web itself and then used to construct the final URL to be crawled.
\item \textbf{Dynamic URLs}. Many platforms allocate content under dynamic URLs. This makes impossible to statically generate a list of addresses to explore. This scenario implies an initial navigation requesting the web platform the URLs and then generating unique URLs that may identify these elements to guarantee their uniqueness in the archive system.

\end{itemize}

As mentioned before, it is difficult to design a generic crawling solution that may be useful in all scenarios~\cite{crescenzi2001roadrunner,Etzioni:2004,Madhavan:2008}. This is due to the fact that many platforms run a massively amount of Javascript code combined with AJAX messaging. Although some works propose methods to deal with automatic form query discovery and interaction, we have not explored these approaches. However, after our experience we can summarize some aspects to be considered during the crawling phase.

\begin{itemize}
\item \textbf{Next links collection}. Many web portals display content obtained after querying a database. This implies that not all the existing content is shown at once, only a small portion. In order to view all the existing contents (or the links that make them accessible), we have click in the next button until no more entries are displayed. In this repetitive operation, one thread can explore each page listing a collection of links while other can navigate through the existing links.

\item \textbf{Not all the elements are displayed}. As mentioned before the results displayed in these systems are simply the result of querying a database. In some cases the result set is split into pages that have to be navigated, in other cases, there is a limit in the result set that makes impossible to completely show the existing results. In these cases, it may not be possible to crawl all the existing results as the system is not disclosing these information. In other cases, the system can be queried which means that a smaller granularity in the queries (days instead of weeks, months instead of years) could be enough in order to find all the occurrences. However, this approach does not guarantee the crawling of all the existing elements in cases where queries using a small granularity might exceed the limits of the result set size.

\item \textbf{Massive utilization of Javascript}. In many cases the utilization of Javascript solutions makes particularly difficult crawling these systems. The most simple solution is to emulate the behaviour of users through Javascript engines. However, this imposes a performance penalty in the crawling side due to the excessive utilization of resources of some browsers and poses an additional problem during the crawler design.

\item \textbf{Request limitations}. In order to avoid denial of service attacks (DDoS) many platforms track requesting machines. Exceeding a certain number of requests may result in a temporal suspension of the service for the user. Every platform is different and only a trial an error approach can disclosure what measures are these platforms using.

\item \textbf{Cookies and sessions}. Many systems employ cookies to store session ids that permit the server to identify the query parameters employed by the user. However, these cookies come with an expiration date that sometimes can cut the platform crawling off.

\item \textbf{Out of service}. Crawled servers may go out of service. Many times crawling these servers will simply return an error HTTP request code. However, many systems return a web page informing about the unavailability of the service. Crawled pages should be checked to contain the kind of data we expect from the crawling approach.

\item \textbf{Content-type}. The content type code returned by the HTTP header is particularly relevant when crawling web pages not encoded using standard ASCII or UTF8. This information can prevent problems during coding/decoding stages. From our experience we can conclude that not all the servers may return the encoding type, these situations require additional analysis in order to avoid future problems.

\end{itemize}

\subsection{Crawled items storage}

By storing the crawled items we can create a data archive that can be used for later purposes. In order to have a traceability of the data, we store it containing the data of the archival and the hash of the item. Every stored item can be described using a JSON file with the following fields:

\begin{lstlisting}[
	basicstyle          = \ttfamily,
	  keywordstyle        = \ttfamily\bfseries,
	  language=json,
	  caption={Example of archived crawled item entry.},
	  label={listing:crawled_item}
]
"archive_item":{
 "_id" : ObjectId("56939e1807cc352b666813ac"),
 "source" : "bg",
 "date" : ISODate("2016-01-11T12:20:40Z"),
 "url" : "www.aop.bg?newver=2&mode=show_doc&doc_id=706665",
 "contentType" : "text/html; charset=windows-1251",
 "file" : ObjectId("56939e1807cc352b666813aa")
}
\end{lstlisting}

The information contained in each field is the following:
\begin{itemize}
\item \textbf{\_id:} Identifier of the item.
\item \textbf{source:} Name of the source where the data comes from.
\item \textbf{date:} Date the data was extracted.
\item \textbf{url:} URL where the data was extracted.
\item \textbf{contentType:} Content type of the HTTP request.
\item \textbf{file:} Object containing the file data.
\end{itemize}

\subsection{Available technologies}

There is currently a large number of available web crawlers in open source projects. It is remarkable the Apache Nutch project~\cite{nutch} that offers a complete framework to design distributed and highly scalable crawlers that easily be connected with other solutions from the Apache environment. However, for smaller scale solutions other frameworks such as Scrapy~\cite{scrapy} are more suitable. Scrapy permits to define concurrent crawlers using Python and provides the developer with a framework that manages simultaneous requests and simplifies the connectivity with Django-based applications.

In order to deal with pages using Javascript the most common approach is to use the Selenium driver~\cite{selenium}. This driver, permits to connect a large number of browsers such as Firefox or Chrome with the crawler and emulate the behaviour of a user performing clicks or typing on top of the current page. However, running a large number of browsers brings a remarkable overhead that can be mitigated using headless Javascript engines such as PhantomJS~\cite{phantomjs}. As a drawback there are some operational limitations when compared with the other full browsers.

\newpage

\section{Structure analysis}
\label{sec:htmlxml}

In our approach we are dealing with structured, semi-structured and unstructured data. The structure analysis component is designed to find the underlying structure of semi-structured and unstructured files and represent this structure in an easily processable format. We have chosen XML as the common format for reasons. (i) It represents hierarchical structures and permits repetitive components (lists). (ii) There are a large number of tools that permit to process XML data. 

The structured data, usually provided in tabular formats, such as Comma Separated Value formats (CSV) or Microsoft Office (or Open Source alternatives) Excel sheets, makes the processing simpler since the number of columns is fixed and the names are provided in the header of the files. 

In the semi-structured category, most of the processed XML files are provided without the XML schema. This makes validation and data extraction more difficult. Moreover, in some cases, the input files were not formatted according to the XML standards (e.g. manually indexed XML elements, elements that can be parsed using regular expressions only, etc.).

The most difficult problem, is transforming the unstructured documents, currently HTML, into structured data. As seen from the state of art (see Section~\ref{sec:relatedwork}), previous approaches deal with this problem by either building regular expressions or XQueries to extract the exact HTML element from the document or build entity relations between different concepts. In the case of public contracts, these approaches are very difficult to use for various reasons. The difficulty of the HTML conversion is increased by the fact that the problem was approached in an unsupervised way, without having any annotated data as input. A preliminary analysis was performed to identify the basic HTML structure of a few documents for each template, but no thorough annotation has been done.

\subsection{The problem of HTML to XML conversion}

In the context of this work, we consider the problem of converting HTML into XML. This transformation is straight forward and not relevant for our purposes. We refer to this transformation as the extraction of the information contained into an HTML document so we can create the corresponding XML file. For example, for the following HTML fragment:

\begin{lstlisting}[
	basicstyle          = \ttfamily,
	keywordstyle        = \ttfamily\bfseries,
	language=xml,
	caption={Example of HTML document to be converted.},
	label={listing:conversion_example_html}
]
...
<span>List of car manufacturers</span>
<ul>
	<li>Name: <b>Audi</b></li>
	<li>Name: <b>Ford</b></li>
	<li>Name: <b>Volkswagen</b></li>
</ul>
...
\end{lstlisting}

The previous example is a simple list of key, value items where the key is \textit{Name} and the value is the car manufacturer. The corresponding XML must reflect the fact that we have a set of repetitive elements which key is \textit{Name}. A possible output is described below:

\begin{lstlisting}[
	basicstyle          = \ttfamily,
	  keywordstyle        = \ttfamily\bfseries,
	  language=xml,
	caption={Example of XML resulting document.},
	label={listing:conversion_example_xml}
]
...
<listOfCars>
	<Name>Audi</Name>
	<Name>Ford</Name>
	<Name>Volkswagen</Name>
</listOfCars>
...
\end{lstlisting}

The conversion process needs to extract all "potentially" useful data and create a common structure, which can be exploited later by various algorithms. In most of the cases, the data is structured into a section or form item (which we call a \textbf{key}) and the content of the key (called the \textbf{value}). The key and the value can be separated by a random sequence of characters (formatted or not). These characters can be either text, HTML formatting tags, new lines or various blank symbols. 

In an ideal situation, the key and the values will be displayed in one of the three cases:

\paragraph*{Case 1:} HTML table structures

\begin{tabular}{|l|l|}
	\hline
	\textbf{Key} & \textbf{Value} \\
	\hline\hline
	Title of the document & \textit{Building a new bridge in Cambridge} \\ \hline
	Description of the document & \textit{Building a new bridge in Cambridge} \\ \hline
	Price & \pounds\textit{1m} \\ \hline
	Contractor & \textit{Some contractor} \\ \hline
\end{tabular}

\paragraph*{Case 2:} Hierarchical contract structures

\begin{listing}
\verb!   !Section 1: Contractual details\\
\verb!      !Section 1.1: Title\\
\verb!      !\textit{Building a new bridge in Cambridge}\\
\verb!      !Section 1.2: Description\\
\verb!      !\textit{Building a new bridge in Cambridge}\\
\verb!   !Section 2: Contractor\\
\verb!      !\textit{Some contractor}
\end{listing}

In this case, the name of the sections are the keys and the content of the section is the value. Moreover, the section numbering is optional.

\paragraph*{Case 3:} Flat contract structure

\begin{listing}
Title of the document : \textit{Building a new bridge in Cambridge} \\
Description of the document = \textit{Building a new bridge in Cambridge} \\
Price\textvisiblespace \pounds\textit{1m} \\
Contractor $\rightarrow$ \textit{Some contractor} \\
\end{listing}

In this example, the key and the value are separated by a fixed set of characters, known at parsing time.

Unfortunately, in most of the existing documents, the three cases are mixed into the same sample, therefore the keys, values and separators are more difficult to identify.

\subsection{Our approach}

According to the scenario described in the previous section, our structure analysis solution has to create the key, value XML file corresponding to a given HTML file. In particular, our approach we are suggesting is based on several empiric observations: 
\begin{itemize}
	\item the search space for the keys is smaller than the values space
	\item the values have to be semantically and visually linked to the key 
	\item the separator can be usually ignored, if the semantic link between the value and the key can be determined without using the separator

\end{itemize}

First, the keys are usually easier to detect than the values and the delimiters between them. These keys are part of a finite set of strings, which are part of the template we are dealing with. Moreover, similar strings can be grouped into similarity sets. We use the concept of synset\footnote{We use the concept of synset a little bit different than the WordNet~\cite{wordnet} and linguistic community does.} without necessarily grouping the elements of a synset by a synonymic relation, but functionally the elements represent the same concept, for our context. Specifically, these elements are plurals of another element of the synset, include determinant forms of sentences, etc. Most of these features can be detected automatically to create the synsets and we employ a fuzzy comparison algorithm, based on a Levenshtein~\cite{Navarro:2001} string similarity, to generate these synsets. 

The synsets are uniquely identified by the key ID and we re-use this to reconstruct the structure (XML) document later into the process. The resulting collection of all synsets creates the dictionary we use in the conversion process. Algorithm \ref{code:baseAlgorithm} presents the basic structure of the key and value matching.

\begin{algorithm}[H]
\caption{Basic structure of the \textbf{(key, value)} matching algorithm}
\label{code:baseAlgorithm}
\begin{algorithmic}
\FORALL{$d \in Documents$}
	\FORALL{$path \in d$}
		\IF[\#using "fuzzy" comparison]{p is a key}
			\STATE previousKey $\leftarrow$ path
		\ELSE[\#path may be a value]
			\IF{similarity(path, previousKey) > similarityThreshold}
				\STATE \COMMENT{\#path is a value related to previous key}
				\STATE storePair(previouskey, path.text)
			\ENDIF
		\ENDIF		
	\ENDFOR
\ENDFOR
\end{algorithmic}
\end{algorithm}

The values have a higher variability than the keys and a dictionary could not be defined. Nevertheless, we exploit the document formatting structure to link the HTML element path of the \textbf{value} to the element path of the previously detected key. This comparison is done with two similarities:
\begin{itemize}
	\item A Levenshtein list similarity, used to compare two HTML element paths with an Edit Distance Algorithm
	\item A longest sequence similarity, used to compare the longest prefix of two HTML element paths, normalized with the length of the compared sequences.
\end{itemize}

In most of our experiments, based on some empiric observations, the second similarity proves to be more effective in matching the key and value.

The element path can be generated in many ways, but we use three ideas for this part:
\begin{itemize}
	\item Using just the tag name of each element in the path, resulting in simpler path, such as: \verb!html\body\div\div\div\p!
	\item Using just the a unique tag name of each element in the path, obtained by uniquely identifying all the HTML elements parsed, such as: \verb!html1\body1\div1\div2\div3\p! or \verb!html1\body1\div34\div36\div39\p!
	\item Using a unique tag name and some of the tag attributes, inspired by the Firefox HTML debugger. This results in a path where the first non-empty identification attribute (\textbf{id}, \textbf{class} or \textbf{name}) is appended to the unique tag name, such as:  \verb!html1\body1\div1.id1\div2.class23\div3.name12\p!. The identification attribute is not necessarily unique and the semantic linking algorithm uses this information.
\end{itemize}

In practice, depending on the HTML style of document, any of these distances can be used, and the choice is made depending on the processing speed in relation to the accuracy of the conversion. The processing speed slowly decreases from the first strategy to the last.

Using these distances and HTML paths to reconstruct the semantic link between the key and value is more robust to template variability and assumes that a selected corpora represents the same information, without assuming that the documents use the same formatting style. 

\subsection{Creating the dictionary}

In the current approach, the creation of the key dictionary is a semi-automatic process, done for each individual language. The automatic part of the algorithm is based on the assumption that the keys will have higher frequencies than the values, given a small sample of the input HTML documents. The sample size depends on the number of HTML document templates identified within the corpora and the variability of the keys. Even though the keys will have higher frequency than the rest of information, there are other static values that need to be manually filtered, such as document type, procedure type, etc. 


The expert can build and enhance the accuracy of the dictionary in a non-supervised approach. This iterative process will permit in a limited number of iterations the quality of the HTML to XML transformation.

\subsection{Flat and hierarchical dictionaries}

In most of the documents processed within the project, even if the document had a hierarchical structure of the sections (keys), most of the leaf sub-section can uniquely be identified by the name. This assumption fits well with most non-repetitive structures as well. 

In some of the collected documents, this assumption was broken by the structure and the name of the sections do not uniquely identify a concept. One, given bellow, is the \textbf{name} identifier:

\begin{listing}
\verb!   !Section 1: The contract details\\
\verb!      !Name\\
\verb!      !\textit{Building a new bridge in Cambridge}\\
\verb!   !Section 2: Contractor\\
\verb!      !Name\\
\verb!      !\textit{Some contractor}
\end{listing}

In this case, we need to create hierarchical keys to represent the information. For now, our approach support hierarchical parsing, but the semi-automatic process for dictionary generation is not supported, therefore the hierarchies have to be done manually.

\subsection{Improving the accuracy of the base algorithms}

In certain cases, the current algorithm does not perform well because of the structure of the data. While most of the data can still be extracted automatically, some of the data fields need to be extracted manually. These manual rules are inserted into the algorithm at various points (loading, pre-processing, post-processing) and the data is mostly re-inserted back into the main loop. We call this process \textit{Data Rescue}.

\paragraph*{Changing the root of the document:} In some cases, the HTML contains some metadata use to display search results or to help the navigation process. Even if the metadata contains the similar data and the fields may have the same name, in most cases these fields do not contain the desired data. The root of the document is changed to the most relevant HTML root element, to improve the accuracy of the overall process.

\paragraph*{Processing elements outside the (key, value) scope:} While most of the elements in this type of data are using a key, value template, described previously, some data elements do not have keys (mainly the title of the document). These elements are manually recovered after the document is loaded, before the processing algorithm starts.

\paragraph*{HTML element rescue:} Most of the elements processed by our algorithm assume that the HTML element path is relatively simple (without creating any restrictions to the length of the path) and the text value is available as a leaf element of the path. In some isolated cases, the HTML code use to display a HTML element is very complex, contains multiple embedded lists, multiple form elements linked by Javascript code to display the same information, etc. In this case, a HTML element rescue is performed manually, the html element code is simplified and inserted back into the algorithm. This technique is employed by only 1 corpora, out of 14.

\newpage

\section{Mapping from XML to Database}
\label{sec:mapping}

We define mapping as the operation that makes possible to identify the correspondence between the elements stored in data model $A$ and data model $B$. The particularities of every model can make extremely difficult the definition of a mapping and/or require of a previous experience with both models. It exists various DSLs (Domain Specific Languages) designed in order to describe the mapping between models. A very well knwon example is XSLT (Extensible Stylesheet Language Transformations)~\cite{xslt} that permits to define transformations between two XML files. However, in many cases mappings can only be defined by experts and these experts may not have any XSLT background.

We reformulate the mapping problem as a knowledge extraction problem where an expert has to define the mapping between two models. In our case, model $A$ (or input model) is an XML file and model $B$ (or target database) is the definition of a database. The transformation process might be particularly complex and it includes some logic operations. Languages such as XSLT support this logic during the transformation process. However, we can simplify the design by assuming the mapping as a collection of XPaths that map the content from the input model into the target model.

\subsection{MITFORD}

In order to simplify the mapping operation and assuming that the expert may not have any kind of IT background we have developed an online tool to maximize the knowledge extraction while minimizing the time the expert has to spend and unifies the development efforts. MITFORD (as we call the tool) permits the expert to iteratively define the mapping between an XML model and a database schema. MITFORD is database and XML file independent making it multi-purpose and reusable in different scenarios. The final output generated from MITFORD is a mapping definition file that can be used in the parsing module (see  Section~\ref{sec:pipeline:parsing}) to populate a database with data from other sources.

\begin{figure}[h]
\centering
\includegraphics[width=\linewidth]{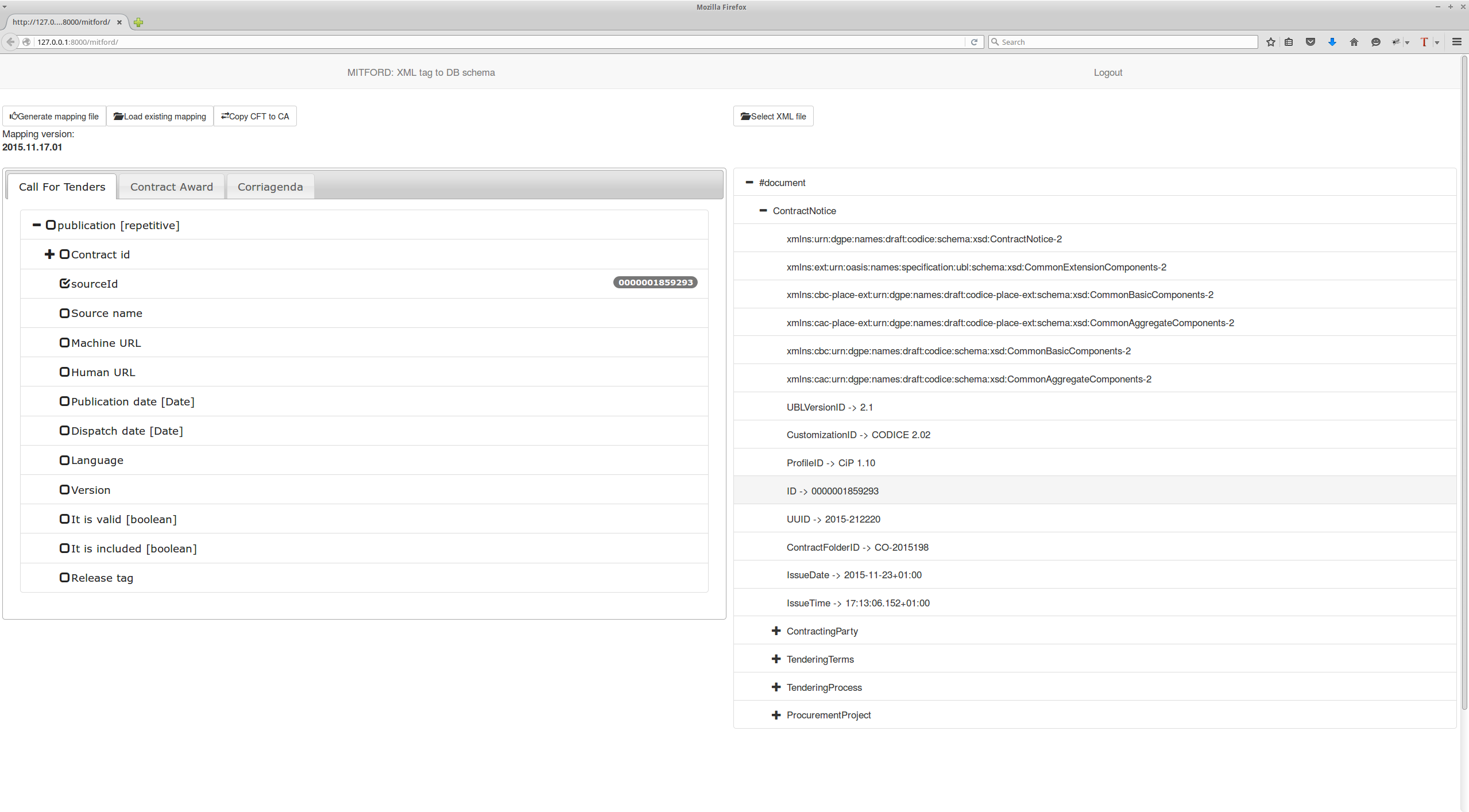}
\caption{Snapshot of MITFORD}
\label{fig:mitford}
\end{figure}

MITFORD is based on an iterative process where the expert load several XML files and define all the mappings she consider relevant for the mapping definition. Once the mapping is finished, the expert can generate the mapping file to be used by in the parsing process. If the mapping is not correct or complete, the operation can be repeated until the final result is achieved.

\subsection{Database schema representation}

MITFORD represents a database schema in a visual frame. The schema is represented using a declarative language that specifies:
\begin{itemize}
	\item \textbf{db}: name of the entity or attribute in the database.
	\item \textbf{repetitive}: boolean field indicating whether the field can appear multiple times or not.
	\item \textbf{text}: text field to display in MITFORD.
	\item \textbf{type}: type of the entry in the database entry model, DateTimeField, TextField, ForeignKey.
	\item \textbf{nodes}: set of other entries contained into the previous one.
\end{itemize}

The database schema representation can be written in a JSON format like the one shown in Listing~\ref{list:database_schema}. In this example \textit{Entity1} has three attributes of type TextField, NullBooleanField and ForeignKey. The attribute3 is a ForeignKey which points to other entity which single attribute (called attribute3-1) has to be filled in. We expect \textit{Entity1} to be repetitive and that is why the attribute \textit{repetitive} is set to \textit{True}.

\begin{lstlisting}[
	basicstyle          = \ttfamily,
	keywordstyle        = \ttfamily\bfseries,
	language=json,
	caption={Database schema representation example},
	label={list:database_schema}
]
{
"db": "Entity1",
"nodes": [
       { "db": "attribute1",
         "repetitive": "false",
         "text": "Name to display attribute1",
         "type": "TextField"
       },
       { "db": "attribute2",
         "repetitive": "false",
         "text": "Name to display attribute2",
         "type": "NullBooleanField"
       },
       { "db": "attribute3",
         "repetitive": "false",
         "text": "Name to display attribute3",
         "type": "ForeignKey",
         "nodes": [
         	{"db": "attribute3-1",
	         "repetitive": "false",
    	     "text": "Name to display attribute3-1",
        	 "type": "NullBooleanField"
         	}
         ]
       },
       ]
"repetitive": "true",
"text": "Estimated value",
"type": "Model"               
}
\end{lstlisting}

\subsection{Mapping file structure}

The structure of the mapping file is a collection of XPaths that permit to identify what elements from the input file are mapped into what elements of the target database schema. For example, assuming the XML file described below:

\begin{lstlisting}[
	basicstyle          = \ttfamily,
	  keywordstyle        = \ttfamily\bfseries,
	  language=xml,
	caption={Input XML example},
	label={list:input_mapping_example}
]
<xml>
	<document>
		<attr1 "name"="document1">document 1</attr1>
	</document>
	<document>
		<attr1 "name"="document2"/>
		<attr2>YES</attr2>
		<child>
			<attr1>Child attribute</attr1>
		</child>
	</document>
	<document>
		<attr1>document 3</attr1>
	</document>
</xml>
\end{lstlisting}

In the previous example, we have a list of documents labelled with  \textit{<document>}. Each document may have a set of attributes and some of them can contain another entity. After finishing the mapping using MITFORD, we come up with the following mapping file:

\begin{lstlisting}[
	basicstyle          = \ttfamily,
	  keywordstyle        = \ttfamily\bfseries,
	  language=json,
	  caption={Database schema representation example},
	label={list:mapping_example}
]
{
    "version": "2015.11.17.01",
    "document": {
	    "attribute1": {
	    	"__xpath__": [
	            "/document/attr1",
                "/document/attr/@name"
            ]
        },
        "attribute2": {
	        "__xpath__": [
    	 	    "/document/attrBoolean"
            ],
            "__conversion__": {
            	"NO":"false"
            }
        },
        "childEntity":{
        	"attribute1": {
	    		"__xpath__": [
	            	"/document/child/attr1",
    	        ]
        	},
        	"__xpath__": [
        		"/document/child"
        	]
        },
        "__xpath__":[
        	"/document"
        ]
    }
}
\end{lstlisting}

The resulting mapping is a JSON file indicating the version of the file and the mapping for the entries in the target database. In the example, we have a \textit{document} entity containing two attributes (\textit{attribute1} and \textit{attribute2}) and a childEntity that contains one attribute. The \textit{\_\_xpath\_\_} element indicates the list of possible XPaths where the data can be found. The extracted data corresponds to the first found entry. If the specified XPath returns several entries, each of the entries is parsed separately following the described structure. In some cases, such as the boolean variables we permit to add some logic to the mapping process by indicating that found variables \textit{NO} should be considered as \textit{false}.

\newpage

\section{Data extraction and storage}
\label{sec:xmlpipeline}

The utilization of a single file format facilitates the utilization of standard parsing libraries. In our case by using XML files we have access to a list of really well tested and efficient libraries. However, there are various issues to be solved:

\begin{itemize}
	\item \textbf{Heterogeneous data formats.} Several download items can be presented in several formats: HTML, zip, tar, tar.gz, etc. A set of consecutive operations have to be carried out in order to prepare the files for later analysis.
	\item \textbf{Inconsistent encoding.} In order to support multiple languages, the utilization of UTF8 encoding is highly recommended. However, we cannot trust the encoding of the original sources. From our experience we can claim that many web servers do not correctly indicate the encoding of the HTTP bodies. This means that it may be necessary to re-encode existing files.
	\item \textbf{Massive number of files.} The amount of files to be processed, and the iterative nature of the problem makes necessary to deal with efficient I/O solutions that can take advantage of multi-core infrastructures.
	\item \textbf{Dependencies among tables.} In the case of E/R platforms the interdependencies between tables may lead to situations of failure due to unfulfilled dependencies.
\end{itemize}

\begin{figure}[H]
	\centering
	\includegraphics[width=\linewidth]{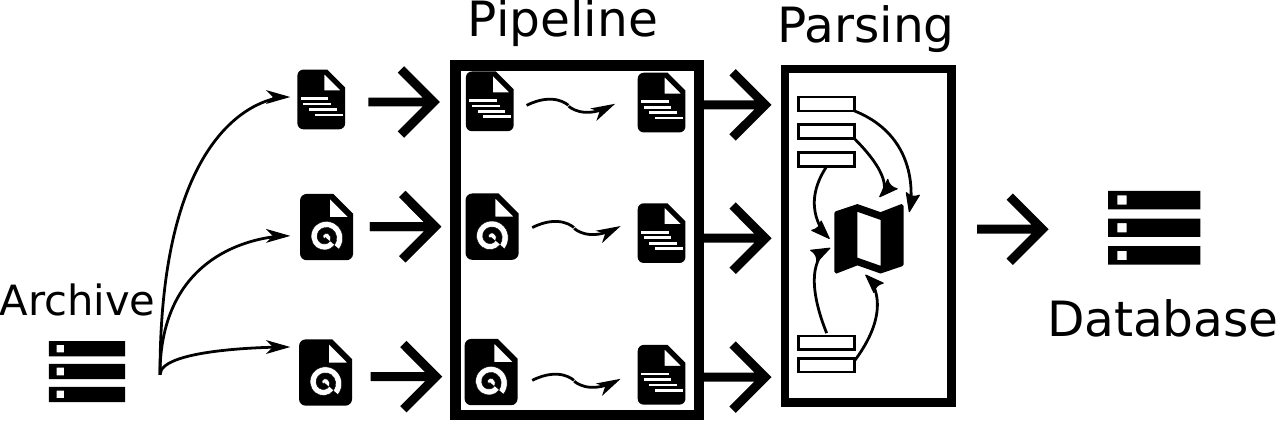}
	\caption{System pipeline}
	\label{fig:pipeline}
\end{figure}

In order to ensure that the input file to the parsing component is homogeneous, this means same format file (XML) and encoding (UTF8), we use a parallel pipeline solution that processes all the input files until obtaining this homogeneous set of files. Figure~\ref{fig:pipeline} illustrates the pipeline process. First the elements are retrieved from the archive. Depending on the initial format, they are sequentially processed applying different transformations until we obtain the final XML file. The files are processed in batches in a multiprocessing environment dispatching several files at the same time. The resulting file after this sequence of transformations can be processed by the parsing module. 

\subsection{Pipeline definition}
The pipeline is defined as a sequence of steps where the input of the next step depends on the execution of the previous one. We can differentiate the processing steps before the file is parsed (preprocessing) and after the file is parsed (post-processing). In our case, preprocessing operations prepare the XML files to be parsed and post-processing basically cleans up the intermediate files generated during the preprocessing. An example of pipeline definition is shown below.

\begin{lstlisting}[
	basicstyle          = \ttfamily,
	  keywordstyle        = \ttfamily\bfseries,
	  language=json,
	  caption={Example of pipeline definition},
	  label={listing:pipeline}
]
PREPROCESS_PIPELINES = {
	"ted":[
		'parsing.pipeline.getFile',
		'parsing.pipeline.uncompress'
	]
}

POSTPROCESS_PIPELINES = {
	"ted":[
		'parsing.pipeline.removeFile',
	]
}

\end{lstlisting}

In this example, we define two operations for the data source called \textit{ted} during the preprocessing. First, we take the corresponding file, then we uncompress this file. In the post-processing, we simply remove all the intermediate generated files.

\subsection{Parsing}
\label{sec:pipeline:parsing}

The parsing system is based on the mapping between the XML input file resulting from the pipeline execution, and the target database schema as explained in Section~\ref{sec:mapping}. Any XML file can be queried using XPaths, although our experience reveals that many input files, even those originally released in XML formats, may not correctly follow an XML design. By correct XML design we mean the following:
\begin{itemize}
	\item Two elements with the same XPATH refer to repetitive elements. In the example below, subElement are different occurrences of an entry.

\begin{lstlisting}[
	basicstyle          = \ttfamily,
	  keywordstyle        = \ttfamily\bfseries,
	  language=xml
]
<elementName>
	<subElement>Something</subElement>
	<subElement>Something else 2nd entry</subElement>
	<subElement>Something else 3rd entry</subElement>
</elementName>
\end{lstlisting}

	\item Repetive elements are correctly formed. In the example below repetitive elements modify their names in order to represent their sequential nature. This is a correctly well-formed file, although the element name ruins the self-description of the file. If we want to query how many \textit{subElement} items do we have we have to use regular expressions to find this elements needing a more complex query.
\begin{listing}
\begin{lstlisting}[
	basicstyle          = \ttfamily,
	  keywordstyle        = \ttfamily\bfseries,
	  language=xml
]
<subElement_1>Something</subElement_1>
<subElement_2>Something else 2nd entry</subElement_2>
<subElement_3>Something else 3rd entry</subElement_3>
\end{lstlisting}
\end{listing}		

	\item Hierarchies are correctly represented. In order to represent a hierarchy we expect to find the subElements contained by the upper elements.
\begin{listing}
\begin{lstlisting}[
	basicstyle          = \ttfamily,
	  keywordstyle        = \ttfamily\bfseries,
	  language=xml
]
<son>
	<son>Son1</son>
	<son>Son2</son>
	<son>
		<son>Son of a son</son>
	</son>
</son>
\end{lstlisting}
\end{listing}
			
Something like the following can be considered as a non-correctly represented XML hierarchy.
\begin{listing}
\begin{lstlisting}[
	basicstyle          = \ttfamily,
	  keywordstyle        = \ttfamily\bfseries,
	  language=xml
]
<son>
	<son_1>Son1</son_1>
	<son_2>Son2</son_2>
	<son_3>
		<son_3_1>Son of a son</son_3_1>
	</son_3>
</son>
\end{lstlisting}
\end{listing}		

\end{itemize}

In order to fulfil the previous assumptions some times it is necessary to pre-process the input files in order to have an adequate input. These operations may require of the additional intervention of a developer. Although, these kinds of situations are the result of bad data modelling policies.

\subsection{Parallelism}

The database load can be a demanding task specially in those cases involving a large number of files, or slow pipeline operations. In order to make the most of the available resources, we use parallelism techniques to improve the performance assuming that several database loads may be required before obtaining the expected results. We apply different techniques:

\begin{itemize}
	\item \textbf{Files.} The input files are grouped into batches that are distributed among the available processors. The size of the files is small enough to avoid any dramatic I/O dependency. The files are processed in a FIFO fashion way, where normally the most time consuming files are parsed at the end.
	
	\item \textbf{Free locks.} In order to avoid processes to lock each other, we use lock-avoiding techniques that permit to run loosely coupled threads. 1) After the information is mapped into the destination database, the information is stored in the local memory. After a fixed amount of records is reached, we execute one single database insertion operation containing all the parsed data. This avoid unnecessary I/O collisions during the parsing. 2) The identifiers for the entries in the database use a sequential identifier decided by the parser. In order to avoid collisions between the identifiers, each processor pre-allocates a range of ids from a common shared id counter. This process avoids concurrent access to id registers.
	
	\item \textbf{XQuery caching.} The parsing process may result into a repetitive set of XQueries that may force existing XML processing libraries to perform similar queries time after time. Caching the results of these queries may dramatically improve the performance of the parsing while paying a small penalty for the storing of these results.
	
	\item \textbf{Avoid entity relationships.} The dependencies between entities in the destination database require of a certain insertion order ($a\rightarrow b \rightarrow c$). This insertion order may not impose a significant problem in environments that make possible disabling this feature. However, certain systems may not permit to disable the integrity checking. In those cases, we provide an order insertion tool that defines the order entities insertion in order to fulfil integrity dependencies.

\end{itemize}

\subsection{Processing}

The processing system works using the mapping file described in Section~\ref{sec:mapping} and the current database schema. The mapping file contains all the information to parse the input data and makes the processing data blind as it only depends on the mapping file definition to perform the transformation. The pseudo-algorithm is described below:

\begin{algorithm}[H]
\caption{Element mapping pseudocode}\label{alg:processing}
\begin{algorithmic}
\REQUIRE mainInstance
\RETURN mainInstance
\STATE numOccurrences$\leftarrow$getOcurrences(xpath)

\FOR{i $\in$ (0,numOccurrences-1)}
	\STATE instance$\leftarrow$instantiateElement()
	\STATE instance.id$\leftarrow$getId()
	\STATE processed$\leftarrow$processEntry(instance,i,parentInstance)
	\IF{processed}
		\STATE addToPool(instance)
		\IF{instance is pointed by other entries}
			\STATE update other entries pointers
		\ENDIF
	\ELSE
		\STATE remove instance
	\ENDIF
\ENDFOR

\end{algorithmic}
\end{algorithm}

Algorithm~\ref{alg:processing} follows the structure defined in the mapping file assuming that each of the entries specified there will be found. Because several entries can be found for the same XPaths (see~\ref{sec:pipeline:parsing}) an iterative filling process is executed. This process instantiates one entry from the target database as described in the mapping file. First, the entry is associated with an identifier used as a primary key. This \textit{floatingId} is used as a control structure independent from the target database. Second, if the entry is finally filled (at least one of its attributes) those entries pointing to this entry are updated accordingly. Finally, the instance is added to the pool of pending instances to be eventually stored into the database. In case the instance was not fulfilled it is discarded.

The \textit{processEntry} function is similar to the algorithm described above. It uses a recursive approach setting the attributes of the instance. In the case another element is found, the element mapping function is called to fill a more complex structure.

\subsubsection{Interdependencies}

We assume in our solution that the target database can be described as a superset where $E_n \rightarrow E_{n-1} \cdots \rightarrow E_{1}\rightarrow E_0$ with $E_i$ the different entities of the database and $\rightarrow$ represents a dependency between two entities. For example, for $E_2 \rightarrow E_1 \rightarrow E_{0}$, $E_2$ points to $E_1$ and this one points to $E_0$. This kind of structures are specially suitable for recursive solutions like the one explained in Algorithm~\ref{alg:processing}. Following a recursive approach, when step $i$ is finished, we can update the pointer to the parent structure created at step $i-1$ and so on. However, this approach is not suitable for designs where $E_2 \rightarrow E_0 \rightarrow E_{1}$ where $E_1$ is created after $E_0$.

\subsubsection{Database insertion}

The dependencies generated by foreign keys in E/R databases, make necessary to stablish an order in the database insertion. This is a problem when storing a large number of entities in a database because many of them may incur into missing dependencies. Some database engines permit to ignore the integrity constraints during insertion, which may be an advantage is after finishing the insertion the integrity is assured. However, this may not occur in all the database engines, or it might be a restricted feature in some drivers.

We assume the target database to have a hierarchical design. This makes possible the definition of a dependency graph following $E_n \rightarrow E_{n-1} \cdots \rightarrow E{1}\rightarrow E_0$. This indicates the insertion order to follow in order to avoid integrity issues. This dependency graph can be automatically generated using the top element in the hierarchy. In our approach we combine this solution with a pool of database entries. This pool is filled until a size limit is reached or no more entries are going to be inserted. The pool contains as many queues as entities represented in the database, and the instantiated entries are queued depending on the corresponding entity. When the scheduler decides to flush out the pool, it follows the dependency graph flushing out every queue in one batch insertion operation. This results into a free-locking and very efficient solution to insert a large number of elements into the target database.

\newpage

\section{Validation}
\label{sec:validation}

The purpose of validation in this project aims to prototype a framework for the evaluation of terminology mapping coverages. As discussed in the previous sections, in this project, an online tool MITFORD has been developed, which allows domain experts to automatically map XPaths into pre-defined database schema. That is, with MITFORD, a domain expert is able to identify a set of designated XPaths in accordance with database schema and generate a set of mapped XPath set for each database field.

However, due to that the number of XML files saved in the data repository is extremely large, and the XPath patterns presented in all files tend to be diverse, it is impossible for a domain expert to explore all available XPaths and map designated ones into database schema. Instead, he/she might resort to sample a number of source files according to some rules, and then parses and maps the corresponding XPaths into database schema. This works under an assumption where the sampling scheme should not lose its generality for representing the whole population. As a consequence, it requires a mechanism to justify to what degree we are confident that the sampled data can be used to represent the whole data population.

In this section, we discuss a framework to evaluate terminology mapping by comparing the rate between mapped and unmapped XPaths. The work is carried out is in two phases. In the first stage, we will evaluate terminology mapping from those files whose original formats are XML. Then, in the next stage, we evaluate terminology mappings from those files whose formats are HTML or CSV but have been converted into XML formats by using the proposed techniques as discussed in the previous sections.

\subsection{Validating XML Mappings}
This subsection discusses validation of terminology mappings from XML based files. XML mappings are based upon comparing the rate between the mapped and the unmapped XPath sets. In the project, in order to address the tasks for collecting data from 35 jurisdictions (34 countries + the EU), crawlers have been developed to automatically crawl source data from data repositories. This enables us, for each jurisdiction, to down a complete set of data files. These files will be used to generate a complete set of XPaths that are mapped into database schema. In general, it is impossible to analyse all XPaths from all files due to extremely large amount of data. Instead, a sampled set is used to perform terminology mapping.

To validation if the sampled data is sufficient enough to represent the data population, we would need to extract all XPaths from all available files and evaluate the proportion between mapped and unmapped XPath set. The framework is prototyped as follows:

\begin{itemize}
   \item Generating a complete set of XPaths from all source files;
   \item Extending XPath set by including attributes to the XPaths; 
   \item Removing XPaths whose values are empty to derive a full non-empty set $S_{xpath}$;
   \item Collecting mapped XPath set from domain experts' mapping set $S_{mapped}$;
   \item Deriving unmapped set by $S_{unmapped}$ = $S_{xpath}$ - $S_{mapped}$.
\end{itemize}

Now the unmapped set $S_{unmapped}$ consists of two parts of information, i.e., 1) XPaths that are irrelevant to the database schema and, 2) XPaths that are relevant to the database schema but are missing from terminology mappings. However, it is difficult to determine which xpaths are terminology related ones. It requires domain experts to manually check the conents. Since the size of $S_{unmapped}$ tends to be large, it would be time expensive on checking XPaths one by one.

In this project, we proposed XPath frequency based sampling scheme for checking XPath terminology mappings. Given an XPath, its frequency on the occurrence across all source files has been calculated. Such a frequency is used as a guidance for sampling XPaths. The higher frequency of an XPath occurs in the statistic figures, the higher chance it will be used to present information. As such, if XPaths with high frequency of occurrence are missing from terminology missing, these XPaths will be assigned with higher priority for sampling.

For the confidence of terminology mapping coverage, we have proposed a scheme to compare the proportion between the size of $S_{mapped}$ and $S_{unmapped}$. If the size of $S_{mapped}$ is much higher than $S_{unmapped}$, it would suggest that the majority of XPaths have been mapped, which leads to a higher confidence for the mapping coverage; otherwise, if the size of $S_{mapped}$ is too low, it indicates that too many XPaths are suffering from terminology missing. As a consequence, we will elaborate to sampling schema to check unmapped set. For the case where the size of $S_{mapped}$ is roughly close to $S_{unmapped}$, we would justify that terminology mappings are on the average level. Still, we need to resort to sampling scheme to further check unmapped set.

\subsection{Validating HTML Mappings}
For the validation of HTML mappings, it works on checking the conversion rate from HTML to XML. Given an HTML file, its contents would be converted into XML based upon pre-defined dictionary. The quality of such conversions is based on how comprehensive the dictionary is defined for the purpose of knowledge extraction. If, by checking the conversion rate, a large amount of information has been ruled out, it indicates the defined dictionary collects very limited terms for conversion mappings. A report will be produced to indicate which parts of information are missing during the process of conversion. This would provide a guidance for the collections of more keywords for the dictionary. Building a comprehensive dictionary for terminology mapping should be treated as an iterative process, i.e., performing conversion and examining missing parts to build a better dictionary, re-performing conversion and then re-examining missing parts. The process terminates until the dictionary is good enough to cover the majority of contents when HTML/XML conversion is performed.

After HTML/XML conversion is completed, the converted XML files can be used for terminology mapping by using the MITFORD tool. Since the dictionary used for HTML/XML conversion is constructed by domain experts, it would be leading to more accurate terminology mappings.

\newpage

\section{Case study: public procurement}
\label{sec:htmlprocurement}

The current economical crisis and outbreaks of corruption in public institutions disclosed by media in Europe have provoked an increasing concerning about the use of public resources. The lack of transparency, the difficulty to access public information and the impossibility to identify the players in the public procurement circus make the analysis of public spending a detective operation. Fortunately, recent platforms are pushing governments to implement transparency measures that include the disclosure of data through the web.

In theory, open data would simplify the task of experts in the analysis of public spending. It would additionally, permit citizens to use this data to improve their knowledge about public spending. However, final solutions are far away from this ideal situation. Ignoring the political aspects involved in the disclosure of public spending data, there are a important number of technical and operational issues:

\begin{itemize}
	\item The lack of a common standard makes impossible to maintain a unique solution that can deal with these data.
	\item Inconsistent (or non-existent) APIs that make extremely difficult to access the data in an organized and consistent way.
	\item Poor definition of data types leading to dirty, malformed and inconsistent data from the very source.
\end{itemize}

The mentioned issues make that every country disclosures their public spending data in a completely different way. Additionally, data access is not designed to be machine-oriented through the use of APIs. This makes necessary the utilization of ad-hoc crawlers in order to extract the information contained in the source webs. This task is time consuming and requires of programmers that can deal with technical aspects such as Javascript, AJAX, pop-ups and other common artefacts used in current web pages.

In this Section, we describe the usage of our web data knowledge extraction methodology in the use case of extracting, transforming and loading public procurement information. In particular, we aim at storing data from a list of European countries and store the data into a unified database schema. This comprehends the acquisition, transformation and storage of the data. We explore the different steps and describe what particular enhancements were designed for this use case. Additionally, we discuss the different aspects to be considered after our experience and existing limitations.

\subsection{Overview}

We initially planned to process the data offered by 35 European countries (37 considering TED and its archive). Each data source corresponds to the web page used by public institutions to offer their existing procurement data to the citizens. We summarize these entries in Table~\ref{table:datasources} with their data formats and reasons in case they were not crawled.
\begin{table}[!ht]
\caption{List of candidate data sources and their corresponding data formats.}
\begin{center}
\begin{tabular}{rcccc}
\hline
\# & Code & Country & Format  & Processed/Notes \\ \hline
0 & oldTed & TED (- 2010) & XML & $\checkmark$ \\ \hline
0 & TED & TED (2011-) & XML & $\checkmark$ \\ \hline
1 & PL & Poland & XML & $\checkmark$ \\ \hline
2 & ES & Spain & XML & $\checkmark$ \\ \hline
3 & GB & UK & XML & $\checkmark$ Data between 2011 and February 2015 \\ \hline
4 & FR & France & HTML & $\checkmark$ \\ \hline
5 & RO & Romania & CSV\&XLS & $\checkmark$ \\ \hline
6 & IT & Italy & XML & $\checkmark$ \\ \hline
7 & BE & Belgium & XML & $\checkmark$  \\ \hline
8 & SK & Slovakia & XML & $\checkmark$  \\ \hline
9 & NL & Netherlands & HTML & $\checkmark$ \\ \hline
10 & NO & Norway & HTML & $\checkmark$ \\ \hline
11 & CZ & Czech & HTML &$\checkmark$  \\ \hline
12 & DK & Denmark & HTML & $\checkmark$ \\ \hline
13 & CH & Switzerland & HTML & $\checkmark$ \\ \hline
14 & PT & Portugal & HTML & $\checkmark$ \\ \hline
15 & EE & Estonia & HTML & $\checkmark$ \\ \hline
16 & BG & Bulgaria & HTML & $\checkmark$ \\ \hline
17 & FI & Finland & HTML & $\checkmark$ \\ \hline
18 & SE & Sweden & HTML & $\checkmark$ \\ \hline
19 & IE & Ireland & HTML & $\checkmark$ \\ \hline
20 & LT & Lithuania & HTML & $\checkmark$ \\ \hline
21 & CY & Cyprus & HTML & $\checkmark$ \\ \hline
22 & HR & Croatia & HTML & $\checkmark$ \\ \hline
23 & HU & Hungary & HTML & $\checkmark$ \\ \hline
24 & DE & Germany & covered by TED & \multicolumn{1}{p{6.5cm}}{TED coverage~~~~~~~$\checkmark$} \\ \hline
25 & LU & Luxembourg & covered by TED & \multicolumn{1}{p{6.5cm}}{TED coverage~~~~~~~$\checkmark$} \\
\hline
26 & LI & Liechtenstein & covered by TED & \multicolumn{1}{p{6.5cm}}{TED coverage~~~~~~~$\checkmark$} \\
\hline\hline
27 & RS & Serbia & HTML & \multicolumn{1}{p{6.5cm}}{Only crawled. Difficult HTML structure} \\ \hline
28 & LV & Latvia & HTML & \multicolumn{1}{p{6.5cm}}{Only crawled. Difficult HTML structure} \\ \hline
29 & AM & Armenia & PDF & \multicolumn{1}{p{6.5cm}}{PDF documents only} \\ \hline
30 & AT & Austria & HTML & \multicolumn{1}{p{6.5cm}}{(Paid) Subscription-based access} \\ \hline
31 & GE & Georgia & PDF & \multicolumn{1}{p{6.5cm}}{PDF documents only}\\ \hline
32 & EL & Greece & HTML & \multicolumn{1}{p{6.5cm}}{Dynamic-generated HTML pages. Difficult website navigation.} \\ \hline
33 & IE & Iceland & HTML & \multicolumn{1}{p{6.5cm}}{(Paid) Subscription-based access} \\ \hline
34 & MT & Malta & PDF & \multicolumn{1}{p{6.5cm}}{PDF documents only} \\ \hline
35 & SI & Slovenia & PDF & \multicolumn{1}{p{6.5cm}}{PDF documents only} \\  \hline
\end{tabular}
\end{center}
\label{table:datasources}
\end{table}

From the original plan we discarded 12 data sources. Liechtenstein, Germany and Luxembourg already publish all their related data using the Tenders Electronic Daily (TED). Armenia, Georgia, Malta and Slovenia are discarded because they use only offer the data in a PDF format. Finally, Iceland and Austria require a fee subscription to access the data in an HTML format.

From the list of processed countries, we observe three groups according to the format they make their data available: XML, HTML and CSV (or XLS). Only Romania uses CSV and XLS files. The remaining countries do not provide alternative data formats, apart from HTML.

The analysis and data processing is challenging for various reasons:
\begin{itemize}
	\item The data is presented in different formats (HTML, XML, DOC, PDF, etc).
	\item The main web portal of each country is different, having multiple internal templates and usually without a common layout.	
	\item The web page design is either very poor or highly dynamic and does not facilitate the data extraction process.
	\item Each country uses the national language for all the documents. If provided, the documents in alternative languages (English mainly) are not complete.
	\item APIs that can be used to recover the document archive are usually not available.
	\item Most of the data is available behind a search form. And it is not possible in many sources to know how many entries are available.
\end{itemize}

We have employed our methodology in order to tackle the data extraction of the information contained in these official documents into a final database design. Apart from the challenges exposed above, we have other operational issues in our use case:
\begin{itemize}
	\item There are no experts available to analyse the data and prepare the extraction methodology.
	\item Without a previous analysis of the data, the number of templates available in each data source, mandatory or optional fields, complexity of extracting data are aspects unknown beforehand.
	\item The development group only understand the following languages: English, French, Spanish, Portuguese, Romanian and Italian.	
	\item The whole process has to carried out under a really tight planning.		
\end{itemize}

This is not an ideal scenario for our methodology as there is no expert who can really drive the extraction process. However, we can show how our approach can really make the difference under the set of previously mentioned constraints.

\subsection{Crawling}

In order to crawl the existing data, we have designed ad-hoc crawlers for each data source. The diversity of the web sources and its variable design made impossible to tackle this task using common reusable functions. We had to design ad-hoc solutions dealing with unfortunate html designs, cookies, foreign languages, unknown server requests encoding and server errors among other issues.

Each source offers data in different ways. Table~\ref{tab:crawl_stats} shows the number of crawled items per source. For the sources without comments, we simply extracted each web page containing contracts. In other cases (see the comments) data were extracted from repositories that usually contained the information in an archived format (compressed files). It is particularly difficult to know the coverage of our crawling we do not have access to the exact number of existing contracts per country. The only reference in most cases is the total number of occurrences returned after querying a search engine. However, this is not usually a reliable number because of query limitations, entries with the same id counted twice, etc.

\begin{table}[!ht]
\caption{Number of crawled items per source.}
\begin{center}
\begin{tabular}{ccrlcc}
\hline
\# & Code & Items & Comments & From & To\\ \hline
0 & oldTed  & 60 &  Monthly backups & 01/01/2011 & 31/12/2015 \\ \hline
0 & TED  & 84 &  Daily packages & 01/04/2011 & 30/01/2016 \\ \hline 
1 & PL  & 1770 &  Daily backup & 01/01/2011 & 29/01/2016 \\ \hline 
2 & ES  & 62901 & & 29/09/2009 & 03/02/2016 \\ \hline
3 & GB  & 49 &  Monthly backups & 01/01/2011 & 31/01/2015 \\ \hline 
4 & FR  & 38053 & & 11/04/2006 & 19/01/2016 \\ \hline 
5 & RO  & 69 &  & 01/04/2010 & 30/06/2015 \\ \hline 
6 & IT  & 24639 &  Manually collected list & 01/01/2015 & 31/01/2016 \\ \hline 
7 & BE  & 26877 &  & 05/11/2012 & 03/02/2016\\ \hline 
8 & SK  & 22258 &  Querying old documents returns errors & 30/12/2016 & 28/04/2014\\ \hline 
9 & NL  & 76215 &  & 01/12/2015 & 29/12/2015 \\ \hline 
10 & NO  & 2725 &  & 03/02/2011 & 18/11/2015\\ \hline 
11 & CZ  & 337919 &  & 01/02/2007 & 30/11/2015\\ \hline 
12 & DK  & 3479 &  & 01/01/2013 & 31/01/2016 \\ \hline 
13 & CH  & 106851 &  & 01/04/2015 & 03/02/2016\\ \hline
14 & PT  & 106851 & & 01/01/2013 & 31/01/2016 \\ \hline
15 & EE  & 24541 &  & 01/01/2014 & 31/01/2016 \\ \hline
16 & BG  & 62109 &  & 01/03/2010 & 30/01/2016\\ \hline
17 & FI  & 4822 &  & 01/01/2015 & 30/01/2016 \\ \hline
18 & SE  & 16 &  Only currently opened notices & &\\ \hline
19 & IE  & 1509 &  & 02/02/2007 & 31/12/2015 \\ \hline 
20 & LT  & 23789 & & 03/10/2008 & 20/01/2016 \\ \hline 
21 & CY  & 8094 & & 01/02/2011 & 31/12/2015\\ \hline 
22 & HR  & 2013 & & 09/04/2013 & 31/12/2015 \\ \hline
23 & HU  & 33880 & Not fully crawled after template change & &\\  \hline 
27 & RS  & 25231 &  Crawled using brute force & &\\ \hline 
28 & LV  & 188713 &  & & \\ \hline 
\end{tabular}
\end{center}
\label{tab:crawl_stats}
\end{table}

We uniquely identified each of the processed items with their URL. However, many pages used dynamic URLs or dynamic frames that do not return a unique URL for every publication. In these cases we create a non-existing URL in order to have a unique reference in the system. All the crawled entries are stored in raw format (exactly using the original encoding) in a Mongo~\cite{mongo} database for its later utilization. Every record has a computed hash to check the integrity of the downloaded data. Data encoding was a major issue during this stage as many servers do not indicate the encoding of the requests. If known, this value is indicated in the record.

\subsection{Structure analysis}

Table \ref{table:datasources} presents all the available data sources, with the document format. Out of these, due to the design of our pipeline, the XML documents are usually fed directly into the mapping process. for the CSV (\&XLS) data, a simple data extractor was written and all the files were converted into XML, allowing us to reuse the mapping tools.

\begin{table}[!ht]
\caption{List of HTML data sources and their degree of difficulty in processing. The index corresponds to the same line in Table \ref{table:datasources}}
\begin{center}
\begin{tabular}{rcccp{8cm}}
\hline
\# & Code & Country & Difficulty  & Notes \\ \hline
4 & FR & France & difficult & Hierarchical Structure, with high variability of the section names \\ \hline
9 & NL & Netherlands & medium & Hierarchical Structure, with semicolon separated sub-fields \\ \hline
10 & NO & Norway & medium & Hierarchical Structure, with semicolon separated sub-fields \\ \hline
11 & CZ & Czech & very difficult & Form-style encoding of the data \\ \hline
12 & DK & Denmark & medium & Hierarchical Structure \\ \hline
13 & CH & Switzerland & medium & Hierarchical Structure. Multiple languages. \\ \hline
14 & PT & Portugal & simple & Table structure \\ \hline
15 & EE & Estonia & medium & Hierarchical Structure \\ \hline
16 & BG & Bulgaria & medium & Hierarchical Structure \\ \hline
17 & FI & Finland & medium & Hierarchical and Table Structure\\ \hline
18 & SE & Sweden & simple & Hierarchical Structure, very few samples \\ \hline
19 & IE & Ireland & simple & Table Structure \\ \hline
20 & LT & Lithuania & simple & Semicolon separated fields \\ \hline
21 & CY & Cyprus & simple & Semicolon separated fields \\ \hline
22 & HR & Croatia & simple & Table Structure \\ \hline
23 & HU & Hungary & very difficult & Mixed: Table, Hierarchical and Semicolon separated structures \\ \hline
\end{tabular}
\end{center}
\label{table:htmlAnalysis}
\end{table}

For the XML data analysis process, a first step of frequency analysis was performed. Table \ref{table:htmlAnalysis} shows an overview of the data structures and their degree of difficulty.

The simple cases usually contain very little data (less than 15 fields) about a contract and an easy to process structure, with a set of clear keywords and separators. The web pages contain a small number of templates, usually defined by the type of document (e.g. Call for Tenders or Contract Award).

In the case of medium difficulty, the countries have a higher variability of the number of templates and keywords. In a usual process, the data processing requires multiple iterations to extract all data, by adjusting various keywords (adding low frequency items) and processing parameters. The quality of the data extracted depends of the number of iterations performed and it usually improves over time.

Processing the French data creates a very difficult configuration due to the structure of the page. The template is structured into various sections, with the names of the sections varying very much from one page to the other, and some title containing even spelling errors. Most of these variations were caught by the automatic synonym detection algorithm, but some had to be merged manually. The number of sections also varies very much from page to page.

In two exceptional cases, the data was very difficult to parse. Hungary has a very high variability, due to non-standardized templates, which makes the processing more difficult than in the French case. This selection of documents was approached with an alternative data processing algorithm.

In the case of Czech data, the documents contain a form-like HTML style, which uses some Javascript mechanisms to select a value in a list, form elements to decorate the value of a field, etc. The HTML parse was not able to retrieve this data automatically, so we had to write a separate set of rules to extract the values from the document fields. In addition, the Czech documents use a hierarchical section structure, with repeating elements, therefore the previous flat processing algorithm was not able to extract the data. A small sample hierarchical model was written for this template, and the whole database was extracted with an alternative approach.

\subsection{Mapping}

In order to define the mapping between the crawled data and the destination database, we have employed an extended version of MITFORD. This version was configured using a database schema for public procurement data. We slightly modified MITFORD to include the idea of call for tenders (cft) and contract awards (ca) as similar but independent data records. This means that every record to be mapped can be classified as a cft or ca, and during mapping we will make a clear distinction between two entities.

For every data source showed in Table~\ref{tab:crawl_stats} we map the XML entries, running a structure analysis in case the original record was described using HTML. The operation is repeated using different files in MITFORD that may represent the different fields that can be found in the original source. This comprehends mapping cft and ca samples until a complete mapping is found.

Additionally, due to the particularities of the public procurement data, we have extended MITFORD to indicate whether fields are repetitive, boolean or dates. From the observation of the initial data, dates can be shown in many different formats. Sometimes the data is split across different fields being elements such as day, month, year and hour split across the document. In other cases, the date is contained into a single string. In the case of boolean entries, we make possible to indicate whether an entry can be translated into false of true. E.g.: \textit{no} is equivalent to false in Spanish, while \textit{nein} is false in German.

During the mapping we observed many issues coming from bad programming practices specially with originally designed XML files. For example, in Poland original files were available in XML format. However, the utilization of repetitive structures were conceptually incorrect. For example, in Listing~\ref{listing:example_pl_file} there is an example of multiple entries described in Polish data. Instead of repeating the same element name, the number of entry is included in the name. This is a bad programming practice that skips the assumptions done by MITFORD.

\begin{lstlisting}[
	basicstyle          = \ttfamily,
	  keywordstyle        = \ttfamily\bfseries,
	  language=xml,
	  caption={Example of pipeline definition},
	  label={listing:example_pl_file}
]
<cpv1c>091300009</cpv1c>
<cpv2c>091320003</cpv2c>
<cpv3c>091341008</cpv3c>
<cpv4c>091330000</cpv4c>
\end{lstlisting}

Fortunately, the utilization of the XQuery language makes possible to design specific queries to solve these situations. This implies the manual modification of the resulting mapping which is not the most convenient solution. In similar cases, what we do is to add specific pipeline operations that transform input XML files into a more reasonable format while maintaining the same information.

\subsection{Parsing}

The parsing task is tightly coupled with the previously explained mapping task. Both parsing and mapping are database dependant in the sense that the same input file will have different mappings depending on the target database design. Variations in the database design will require modifications in the obtained mapping. Although there is a significant impact in the modification of the database design, this is smaller compared with a ``hard-coded`` solutions.

In order to simplify the administration of the database and reduce the development time, we described the database using the Django Object-Relational Mapping (ORM)~\cite{django_orm}. In this way, we abstract away the underlying database engine and take advantage of all the existing functions available in Django. Like in any database engine the fields contained in each record are typed. This means, that during the description of the database these types have to be indicated accordingly. However, our data exploration during the mapping reveals that most of the fields to be parsed are not typed. If the collected data is not correctly typed in origin, this means that during parsing we cannot be sure of finding integer, float or data types. To fix these problems we need to design filters to convert originally dirty data into the expected values. However, in order to do this we need to collect all the possible values and then determine how to transform them. Although this task can be included as a consecutive step of our methodology, we postpone this task for posterior work. We simplify the database by considering all the fields to be text typed.

The size of the generated databases varies depending on the amount of available data in the corresponding sources and the thoroughness of the mapping. Figure~\ref{fig:db_sizes} shows the size in megabytes for every generated database. The showed values correspond to the total size of the database after loaded in a PostgreSQL engine. This means that these sizes include data and metadata used by the database (indices, control sequences, etc.). There is large difference among the generated databases (note the logarithmic scale). The largest database is TED with 20 GBytes, followed by Poland with 13 GBytes. Only five databases require more than one gigabyte after storage \textit{ted}, \textit{pl}, \textit{ro}, \textit{it} and \textit{oldTed} with 20, 13, 2.5, 1.6 and 1.3 GBytes respectively. The remaining databases go from 299 MBytes (\textit{nl}) to 14 MBytes (\textit{ie}).

	\begin{minipage}{0.48\linewidth}
			\includegraphics[width=\linewidth]{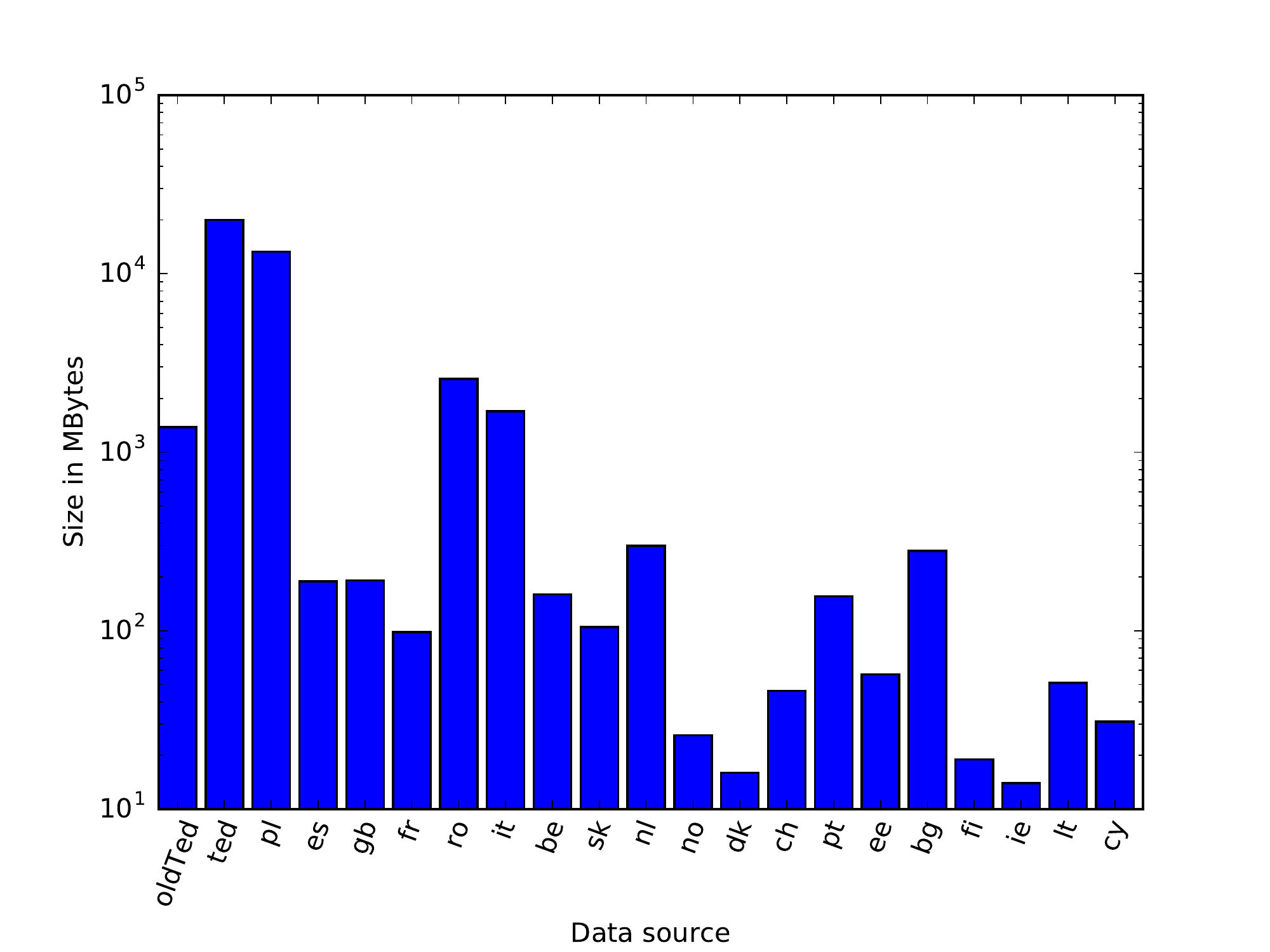}
			\captionof{figure}{Size in MBytes for the resulting databases generated after the parsing process}
		\label{fig:db_sizes}
	\end{minipage}
	\begin{minipage}{0.48\linewidth}
			\includegraphics[width=\linewidth]{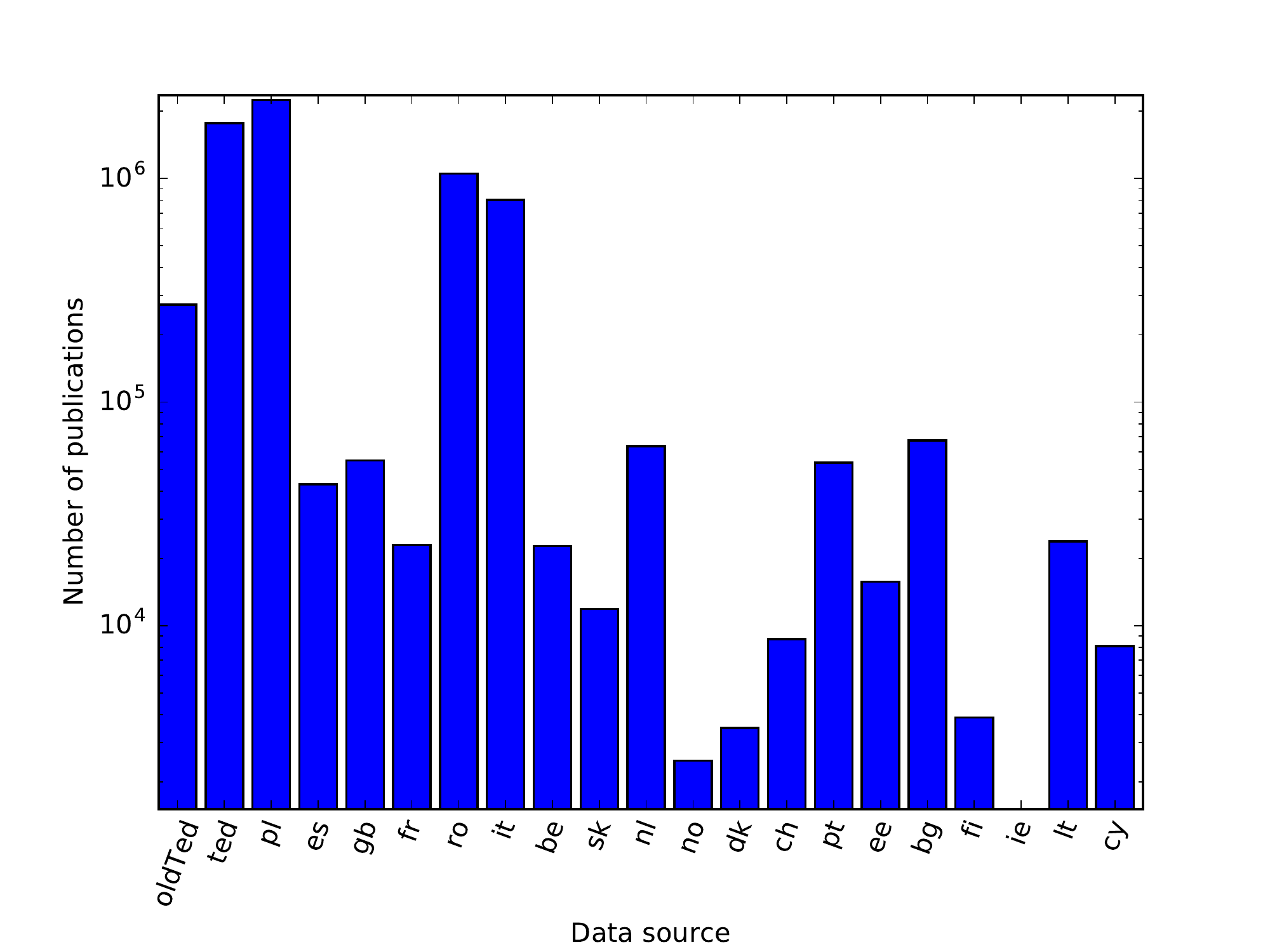}
			\captionof{figure}{Number of publications stored per database. Note the logarithmic scale.}
			\label{fig:db_entries}	
	\end{minipage}


The size of every database depends on the amount of information extracted from the original data source. In some cases, there is an important amount of textual data to be stored, in other cases the mapping is more extensive or the amount of available entries varies depending on the data source. Figure~\ref{fig:db_entries} shows the number of entries per database. As previously observed, there is a large variability in the number of publications we can find. The five most populated databases are the five largest ones. However, the raking changes with \textit{pl}, \textit{ted}, \textit{ro}, \textit{it} and \textit{oldTed} with 2.2M, 1.7M, 1M, 800K and 272K entries respectively.

\begin{figure}[!h]
\centering
\includegraphics[width=0.7\linewidth]{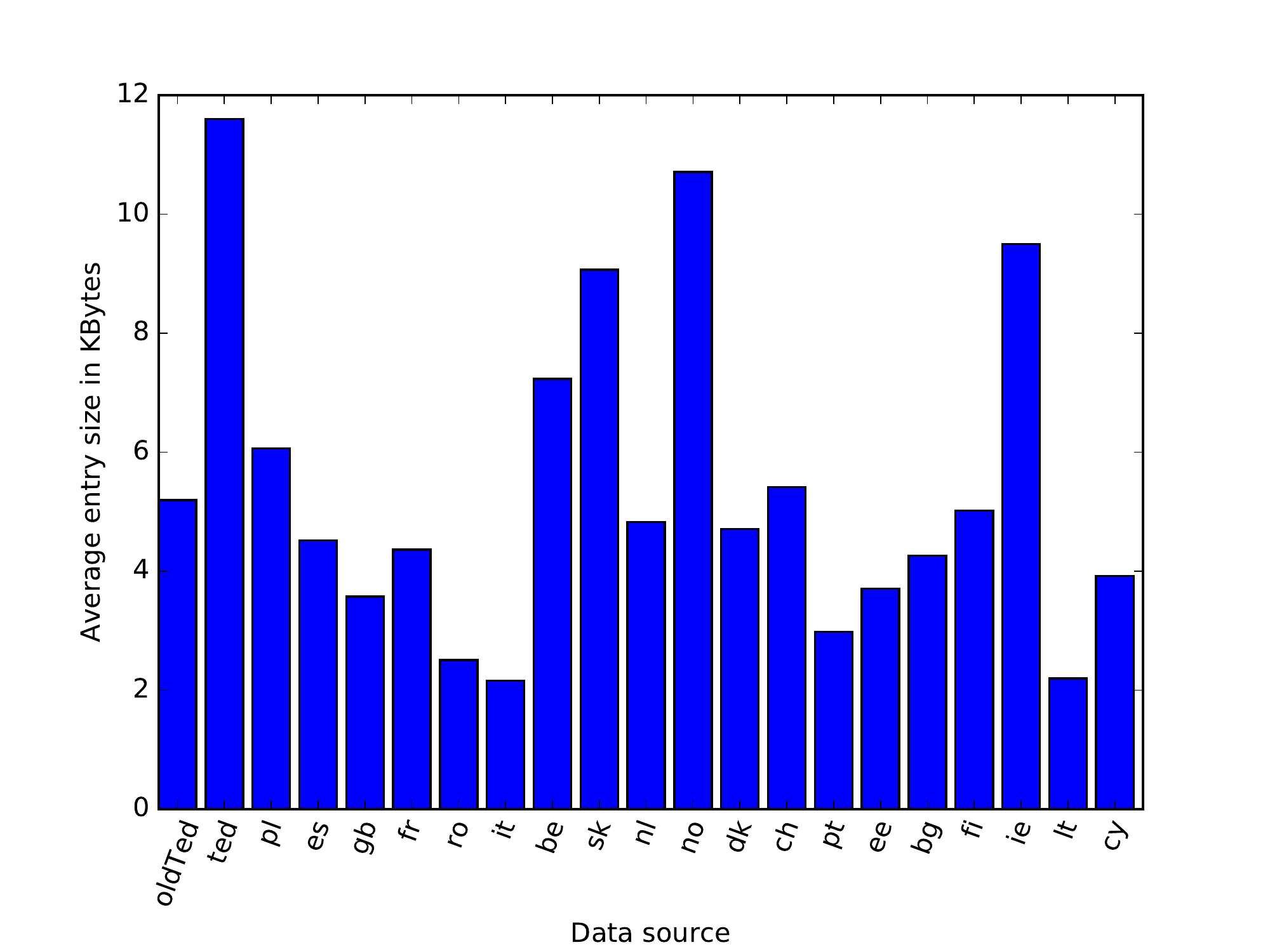}
\caption{Average size in KBytes per entry}
\label{fig:db_avg_size}
\end{figure}

If we compute the average size per entry dividing the the database size by the number of entries (see Figure~\ref{fig:db_avg_size}) we observe a change in the distribution. Most of the databases remain under 6 KBytes. And now the top five largest average entries are for \textit{ted} (11.6 KBytes), \textit{no} (10.7 KBytes), \textit{ie} (9.5 KBytes), \textit{sk} (9 KBytes) and \textit{hr} (8.6 KBytes). This result may indicate a larger textual contain in the records for these databases.

A more detailed analysis of the distribution of the entries per table and database reveals a great unbalance in the number of records allocated per table. We define the table occupation ratio as
\begin{equation}
r_i=\dfrac{|e_i|}{\sum_{\forall i \in T}|e_i|}
\end{equation}
with $T$ the set of used tables and $e_i$ the set of entries in table $i$ with $i \in T$. The ratio oscillates between 0 and 1. Low values indicate a low number of records from the total collection of entries in the database contained in the table. In an ideally balanced database, all the occupation ratios should be $1/|T|$. This is, the same number of records per table. However, this is an ideal scenario extremely difficult to achieve in real world models. The skewness of the distribution of $r_i$ can reveal possible database design failures and facilitate the data analysis.

\begin{figure}[!ht]
\centering
\includegraphics[width=\linewidth]{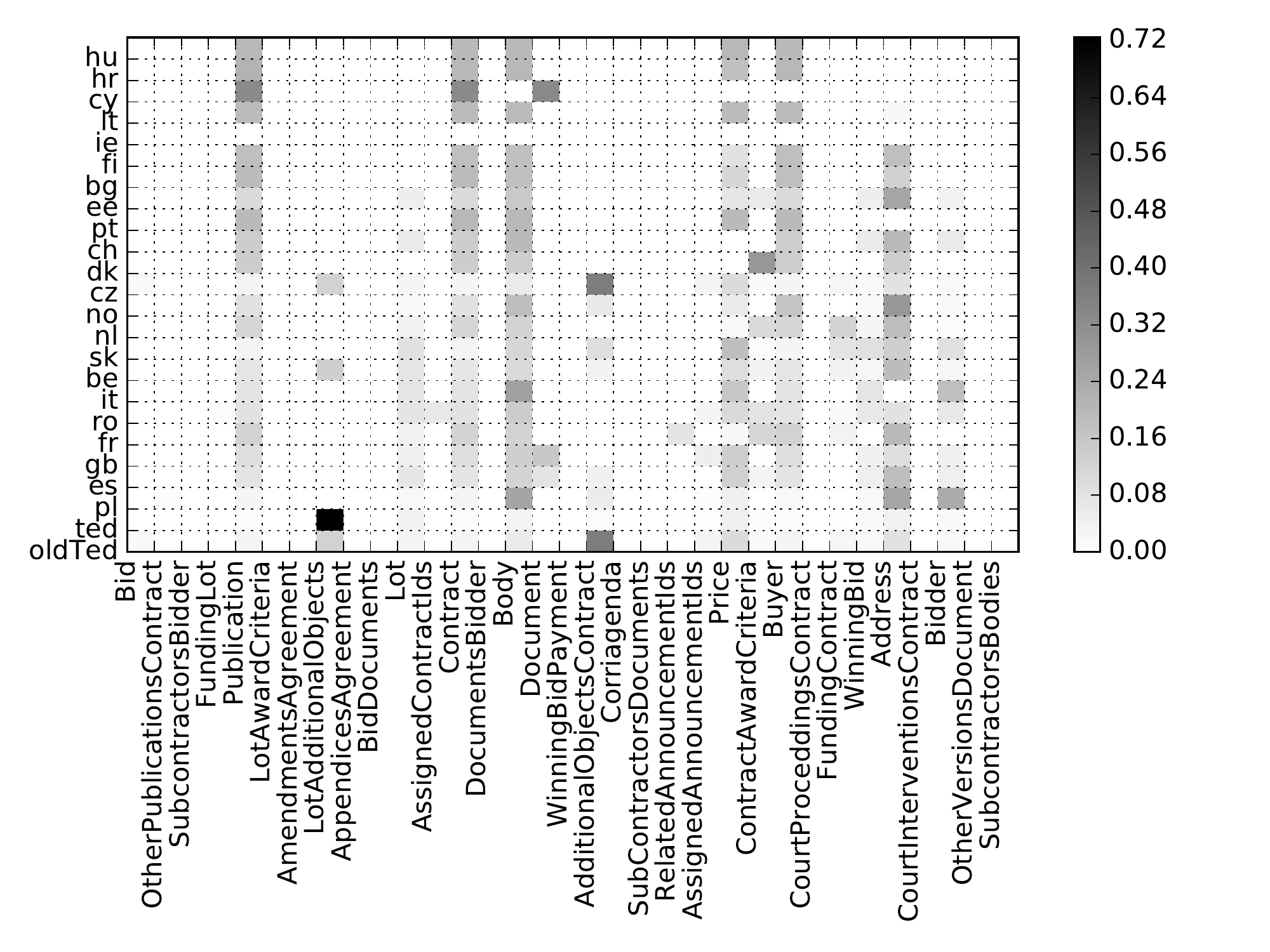}
\caption{$r_i$ per database.}
\label{fig:db_distribution_heatmap}
\end{figure}

Figure~\ref{fig:db_distribution_heatmap} shows the $r_i$ values computed for the databases and the 33 tables allocated in each database. White cells indicate a lack of entries and dark cells indicate a larger number. We observe that most of the tables are empty or almost empty for all the databases. Most of the information is concentrated in \textit{Address}, \textit{Body}, \textit{LotAdditionalObjects} and \textit{Price} tables. The most surprising case is \textit{LotAdditionalObjects} table in \textit{ted} database with $e_i=0.72$. This is due to the utilization of a large number of object descriptors per lot. This analysis indicates that most of the tables are never populated and in some cases only with small amounts of data compared with the size of other tables. The lack of entries in some of the tables may indicate a poor mapping or simply that the information is not available at the origin. In our case, we cannot state any of these situations. The lack of a domain expert makes difficult to have a clear picture of the data available in the data sources and if whether the data regarding these tables is available or not.

\subsection{Periodically maintained pipeline process}

The data sources we crawled are updated on different periods of time (daily, monthly, etc.). The pipeline mentioned above can be run periodically and only new found elements will be processed and stored. Normally, a programmed script can run the crawlers with different periods of time according to the update period of each data source. Then, running the corresponding pipeline will add new found elements to the database.

\subsection{Validation}

Validations were carried out via evaluating terminology mappings from all jurisdictions one by one. As shown in Table \ref{table:datasources}, total 25 jurisdictions have been considered for data collection and data parsing. As such, data from these 25 jurisdictions will be mapped and validated respectively. To save the page limitations in this report, in the following discussions, we will use UK (XML based) and Denmark (HTML based) as two illustrative examples.

For the evaluation of UK terminology mappings, we would first generate a complete set of non-empty XPaths. These XPaths are saved in a JSON file with the format shown as:

\begin{lstlisting}[
	basicstyle          = \ttfamily,
	  keywordstyle        = \ttfamily\bfseries,
	  language=json,
	  caption={Example of XPATH set used for validation.},
	  label={listing:validation_example}
]
{
  "SYSTEM/SOURCE_SYSTEM/PUBLISHED_ON_SOURCE_DATE/DAY": [
    56974,
    [
      "notices_2011_02.xml",
      "notices_2011_11.xml"
    ]
  ],
  "FD_CONTRACT/COMPLEMENTARY_INFORMATION_CONTRACT_NOTICE/
         RECURRENT_PROCUREMENT": [
    23,
    [
      "notices_2014_11.xml",
      "notices_2014_12.xml"
    ]
  ],
  ...
}
\end{lstlisting}

This shows XPath SYSTEM/SOURCE\_SYSTEM/PUBLISHED\_ON\_SOURCE\_DATE/DAY occurs 56974 times across all contract files, while FD\_CONTRACT/COMPLEMENTARY\_ INFORMATION\_CONTRACT\_NOTICE/RECURRENT\_PROCUREMENT 23 times. For each XPath, the names of two contracts that contain the XPath are included.

Such statistics figures show that the first XPath owns a much higher occurrence frequency which entitles it a higher priority for sampling; while, the second has a much lower occurrence frequency and, as a consequence, is assigned with a lower priority for sampling. However, by presenting information of XPaths with lower occurrence frequency, we still intend to provide domain experts an alternative to justify whether these low frequency occurrence XPaths are of any value for database terminology mappings.

Terminology mappings from UK data have already been performed by domain experts. A set of mapped XPaths is saved in a JSON file with the format as shown in \ref{list:input_mapping_example}. Based upon such terminology mappings and the generated complete set of non-empty XPaths, we can derive a set of XPaths that is not mapped into a database schema. Now, according to the occurrence of frequency, we sampled a set of XPaths with higher priority and let domain experts to perform further terminology mappings. By completing this work, we derive two subsets of the XPaths one of which is related to database schema while the other irrelevant. We then included the newly mapped XPaths into the mapping file and repeated sampling operation from the remaining subset of the unmapped XPath set. Such an operation is repeated until the domain expert reaches a degree where he/she agrees that the collected terminology mappings are sufficient for data parsing, and the parsed data are considered to be good enough for any further research activities.

From the experimental results extracted from mapping and validating UK data, we have noticed that, with such an iterative amending scheme, more XPaths can be mapped into database schema in an efficient way. Meanwhile, the proposed framework not only discusses a way to analyse XPaths with high occurrence frequency, but also provides a meaning for domain experts to disclose XPaths with low occurrence frequency. Due to their low occurrence frequency, these XPaths tend to be missed out in the domain experts sampling stage, which could lead to terminology mappings incomplete.

For the studies of validating terminology mappings from Denmark data (HTML based), the validation works on checking the conversion rate from HTML to XML. Initially, domain experts will develop a dictionary which contains a set of keywords used for HTML/XML conversion. Such a dictionary is incomplete, leading to a lot of information being missed. By checking the conversion rate, a report is produced to explain which parts of contents in the HTML files are missed out. An example is illustrated below where domain experts defined a keyword (labelled as K185) for mapping authority based conversion. By checking the converted contents, it was reported that some data from authority were missed from terminology mappings. Based upon such a report, domain experts modified the definition of K185 which makes the terminology mappings more comprehensive.

\begin{lstlisting}[
	basicstyle          = \ttfamily,
	  keywordstyle        = \ttfamily\bfseries,
	  language=json,
	  caption={Example of document used for validation.},
	  label={listing:validation_example_document},
	  extendedchars=true,
	  inputencoding=utf8,
	   escapeinside={\%*}{*)}
]
-----------------------------------------------------------
Del I: Ordregivende myndighed
I.1) Navn, adresser og kontaktpunkt(er)
%*Københavns Kommune - v. Økonomiforvaltningen*)
%*Center for Økonomi*)
Maj Calmer Kristensen
1599
%*København V *)
DANMARK
+45 29664193
b20b@okf.kk.dk

%%%%
<K185>:  Missing
   %*Authority Body : Københavns Kommune*)
   %*- v. Økonomiforvaltningen, (City of Copenhagen*)
   - v. Finance Administration), telephone, email and URL.
-----------------------------------------------------------
\end{lstlisting}

The above process iterated for a number of times until the dictionary is constructed in a way where the majority of data in the HTML files have been covered in the converted XML. From the real terminology mapping work, it can also be concluded such an iterative re-enforcement learning process works well when Denmark data are converted and mapped into database schema.

\subsection{Discussion}

Extracting knowledge from the public procurement web sites enumerated in Table~\ref{table:datasources} is a complex task that demands a large number of working hours. We believe that our methodological approach makes possible to split the tasks involved in the process that goes from the data collection to the data storage. A proof of this is that we manage to build a large set of databases in a short period of time without any expert guidance. Our experience strengthen the idea of using methodologies to assure the success of data extraction projects. There are several aspects to take into consideration in the expert-driven mapping.

\begin{itemize}
	\item \textbf{Know your data in advance}. Mapping, parsing and transforming data is a not an easy task. A prior expert knowledge of the data to be processed is crucial to be successful. This knowledge must comprehend the semantics of the data, and how these is presented. This is, format (XML, HTML, HDF5, etc.), how the data is structured (repetitive structures, boolean fields, hierarchical design, etc.), mandatory and optional fields, particularities of the data, etc.
	\item \textbf{Determine the amount of data to be processed}. Storage solutions must be designed depending on the amount of data we want to store and the purpose of our data. Obviously, deploying regular database engines such as MySQL or PostgreSQL imply more resources than running small solutions such as liteSQL and may serve the same purpose. On the other hand, large deployments may demand distributed storage solutions.
	\item \textbf{Minimalistic design}. In the presented use case, the database schema we have used aims to capture the semantics contained in all the data sources. However, the resulting design is a really large design with hundreds of variables that are not populated most of the time. The semantic overlapping between the different data sources reveals that many of the fields in the database are redundant or unnecessary. This problem is related with the first point of our discussion, as a better understanding of the data would have made possible to reduce the complexity of this design and therefore, the complexity of the mapping and parsing.
	\item \textbf{Make the expert work}. The idea of the mapping makes possible to use the experts as a task force in the project. They can use their knowledge to design a mapping strategy and check whether the results are the expected ones or not. Most of the process should be guided by the expert in an iterative circle of mapping, parsing and checking. In the best scenario the expert should be isolated from any programming aspect and should be able to do his job without interfering with any developer. The comments from the experts can trigger certain changes in the design of specific solutions that may simplify the mapping task.
	
	\item \textbf{Preparing data is an iterative operation.} The data preparing process is an iterative process. Depending on the prior knowledge the number of iterations may vary among data sources. In this particular use case, we couldn't assure the final data type of the database elements. This may make not possible data mining operations based on numbers of date times for example. However, the data remains ready for further cleaning and linking following the already mentioned iterative approach.
			
\end{itemize}

\subsection*{Annex: Database design}
In order to facilitate the database design we employ the Django ORM. This abstraction model permits to define objects and their relationships with other objects using objects in Python. Then these objects and their relationships are mapped onto tables independently of the employed database. Tasks such as database management (tables creation, deletion, modification, etc.), data type changes, addition of new elements, etc. can be automatically carried out by DJango.

We describe the database design employed in this use case. The following tables describe each of the entities in the database and their associated variables. Each line describes the variable name, the type and the foreign table in case of \textit{ForeignKey} types. A more detailed description of the database entries can be found in the code itself.

\begin{longtable}{ p{.40\textwidth}  p{.15\textwidth}  p{.15\textwidth}} 
\multicolumn{3}{c}{\textbf{Publication}}\\
	\hline
	Variable & Type & Foreign table\\
	\hline
    contract  &  ForeignKey & Contract \\ 
    sourceId  &  Text &  \\ 
    sourceName  &  Text &  \\ 
    machineURL  &  Text &  \\ 
    humanURL  &  Text &  \\ 
    publicationDate  &  Text &  \\ 
    dispatchDate  &  Text &  \\ 
    language  &  Text &  \\ 
    version  &  Text &  \\ 
	valid & Boolean & \\ 
    included  &  Boolean &  \\ 
    releaseTag  &  Text &  \\ 
    \hline
    \\
\multicolumn{3}{c}{\textbf{Contract}} \\
	\hline
	Variable & Type & Foreign table\\
	\hline
    id  &  Integer &  \\ 
    assignedContractId  &  Text &  \\ 
    announcementType  &  Text &  \\ 
    country  &  Text &  \\ 
    title  &  Text &  \\ 
    titleEnglish  &  Text &  \\ 
    procedureType  &  Text &  \\ 
    nationalProcedureType  &  Text &  \\ 
    description  &  Text &  \\ 
    descriptionEnglish  &  Text &  \\ 
    bidderLimit  &  Text &  \\ 
    estimatedValue  &  ForeignKey & Price \\ 
    finalValue  &  ForeignKey & Price \\ 
    type  &  Text &  \\ 
    size  &  Text &  \\ 
    callForTendersPublicationDate  &  Text &  \\ 
    bidDeadlineDate  &  Text &  \\ 
    docDeadlineDate  &  Text &  \\ 
    documentsPayable  &  Boolean &  \\ 
    documentsPrice  &  ForeignKey & Price \\ 
    documentsLocation  &  ForeignKey & Address \\ 
    contractAwardNoticePublicationDate  &  Text &  \\ 
    lastUpdate  &  Text &  \\ 
    awardByGroupBuyers  &  Boolean &  \\ 
    isCentralPurchase  &  Boolean &  \\ 
    buyer  &  ForeignKey & Buyer \\ 
    onBehalfOf  &  ForeignKey & Buyer \\ 
    parentContractAwardId  &  Text &  \\ 
    parentContractAwardDate  &  Text &  \\ 
    parentContractPublicationUrl  &  Text &  \\ 
    administrator  &  ForeignKey & Body \\ 
    supervisor  &  ForeignKey & Body \\ 
    specificationsCreator  &  ForeignKey & Body \\ 
    cftVersion  &  Text &  \\ 
    caVersion  &  Text &  \\ 
    npwpReason  &  Text &  \\ 
    variantsAccepted  &  Boolean &  \\ 
    deposits  &  Text &  \\ 
    personalRequirements  &  Text &  \\ 
    economicRequirements  &  Text &  \\ 
    technicalRequirements  &  Text &  \\ 
    priorNotification  &  Boolean &  \\ 
    correction  &  Boolean &  \\ 
    callForTenders  &  Boolean &  \\ 
    contractAward  &  Boolean &  \\ 
    appealBodyName  &  Text &  \\ 
    mediationBodyName  &  Text &  \\ 
    other  &  Text &  \\ 
    coveredByGPA  &  Boolean &  \\ 
    frameworkByAgreement  &  Boolean &  \\ 
    addressOfPerformance  &  ForeignKey & Address \\ 
    isDPS  &  Boolean &  \\ 
    estimatedStartDate  &  Text &  \\ 
    estimatedCompletionDate  &  Text &  \\ 
    awardDecisionDate  &  Text &  \\ 
    contractSignatureDate  &  Text &  \\ 
    electronicAuctionUsed  &  Boolean &  \\ 
    mainObject  &  Text &  \\ 
    \\
\multicolumn{3}{c}{\textbf{CourtProceedingsContract}} \\
	\hline
	Variable & Type & Foreign table\\
	\hline
    id  &  Integer &  \\ 
    contract & ForeignKey & Contract \\ 
    url & Text &  \\ 
\\
\multicolumn{3}{c}{\textbf{CourtInterventionsContract}} \\
	\hline
	Variable & Type & Foreign table\\
	\hline
    id  &  Integer &  \\ 
    contract & ForeignKey & Contract \\ 
    url & Text &  \\ 
\\
\multicolumn{3}{c}{\textbf{AdditionalObjedtsContract}} \\
	\hline
	Variable & Type & Foreign table\\
	\hline
    id  &  Integer &  \\ 
    contract  &  ForeignKey & Contradt \\ 
    cpv  &  Text &  \\ 
\\
\multicolumn{3}{c}{\textbf{FundingContract}} \\
	\hline
	Variable & Type & Foreign table\\
	\hline
    id  &  Integer &  \\ 
    contract  &  ForeignKey & Contract \\ 
    source  &  Text &  \\ 
    euFund  &  Boolean &  \\ 
    programme  &  Text &  \\ 
    amount  &  ForeignKey &  \\ 
    proportion  &  Text &  \\ 
\\
\multicolumn{3}{c}{\textbf{ContractAwardCriteria}} \\
	\hline
	Variable & Type & Foreign table\\
	\hline
    id  &  Integer &  \\ 
    contract  &  ForeignKey & Contract \\ 
    name  &  Text &  \\ 
    weight  &  Text &  \\ 
    description  &  Text &  \\ 
    priceRelated  &  Boolean &  \\ 
\\
\multicolumn{3}{c}{\textbf{OtherPublicationsContract}} \\
	\hline
	Variable & Type & Foreign table\\
	\hline
    id  &  Integer &  \\ 
    contract  &  ForeignKey & Contract \\ 
    publication  &  ForeignKey & Publication \\ 
\\
\multicolumn{3}{c}{\textbf{Address}}\\
	\hline
	Variable & Type & Foreign table\\
	\hline
    id  &  Integer &  \\ 
    rawAddress  &  Text &  \\ 
    street  &  Text &  \\ 
    city  &  Text &  \\ 
    country  &  Text &  \\ 
    postcode  &  Text &  \\ 
    nuts  &  Text &  \\ 
\\
\multicolumn{3}{c}{\textbf{Price}} \\
	\hline
	Variable & Type & Foreign table\\
	\hline
    id  &  Integer &  \\ 
    netAmount  &  Text &  \\ 
    vat  &  Text &  \\ 
    currency  &  Text &  \\ 
    netAmountEur  &  Text &  \\ 
    unitPrice  &  Text &  \\ 
    unitNumber  &  Text &  \\ 
    unitType  &  Text &  \\ 
    minPrice  &  Text &  \\ 
    maxPrice  &  Text &  \\ 
\\
\multicolumn{3}{c}{\textbf{LotAwardCriteria}} \\
	\hline
	Variable & Type & Foreign table\\
	\hline
    id  &  Integer &  \\ 
    lot  &  ForeignKey & Lot \\ 
    name  &  Text &  \\ 
    weight  &  Text &  \\ 
    description  &  Text &  \\ 
    priceRelated  &  Boolean &  \\ 
\\
\multicolumn{3}{c}{\textbf{Body}} \\
	\hline
	Variable & Type & Foreign table\\
	\hline
    id  &  Integer &  \\ 
    bodyId  &  Text &  \\ 
    idType  &  Text &  \\ 
    idScope  &  Text &  \\ 
    bodyName  &  Text &  \\ 
    bodyAddress  &  ForeignKey & Address \\ 
    email  &  Text &  \\ 
    url  &  Text &  \\ 
    contactPoint  &  Text &  \\ 
    contactName  &  Text &  \\ 
    contactPhone  &  Text &  \\ 
\\
\multicolumn{3}{c}{\textbf{Lot}} \\
	\hline
	Variable & Type & Foreign table\\
	\hline
    id  &  Integer &  \\ 
    contractId  &  ForeignKey & Contract \\ 
    assignedContractId  &  Text &  \\ 
    lotNumber  &  Text &  \\ 
    lotTitle  &  Text &  \\ 
    lotTitleEnglish  &  Text &  \\ 
    description  &  Text &  \\ 
    descriptionEnglish  &  Text &  \\ 
    lotStatus  &  Text &  \\ 
    mainObject  &  Text &  \\ 
    estimatedValue  &  ForeignKey & Price \\ 
    finalValue  &  ForeignKey & Price \\ 
    addressOfPerformance  &  ForeignKey & Address \\ 
    estimatedStartDate  &  Text &  \\ 
    estimatedCompletionDate  &  Text &  \\ 
    awardDecisionDate  &  Text &  \\ 
    contractSignatureDate  &  Text &  \\ 
    completionDate  &  Text &  \\ 
    cancellationDate  &  Text &  \\ 
    cancellationReason  &  Text &  \\ 
    electronicAuctionUsed  &  Boolean &  \\ 
    frameworkAgreement  &  Boolean &  \\ 
    estimatedWinnersInFA  &  Text &  \\ 
    isDPS  &  Boolean &  \\ 
    coveredByGPA  &  Boolean &  \\ 
    eligibilityCritera  &  Text &  \\ 
    winningBid  &  ForeignKey & WinningBid \\ 
    bidsCount  &  Text &  \\ 
    validBidsCount  &  Text &  \\ 
    electronicBidsCount  &  Text &  \\ 
\\
\multicolumn{3}{c}{\textbf{FundingLot}} \\
	\hline
	Variable & Type & Foreign table\\
	\hline
    id  &  Integer &  \\ 
    lot & ForeignKey & Lot \\ 
    source  &  Text &  \\ 
    euFund  &  Boolean &  \\ 
    programme  &  Text &  \\ 
    amount  &  ForeignKey & Price \\ 
    proportion  &  Text &  \\ 
\\
\multicolumn{3}{c}{\textbf{LotAdditionalObjects}} \\
	\hline
	Variable & Type & Foreign table\\
	\hline
    id  &  Integer &  \\ 
    lot & ForeignKey & Lot \\ 
    cpv  &  Text &  \\ 
\\
\multicolumn{3}{c}{\textbf{WinningBid}} \\
	\hline
	Variable & Type & Foreign table\\
	\hline
    id  &  Integer &  \\ 
    wasInRequestedQuality  &  Boolean &  \\ 
    wasFinishedOnTime  &  Boolean &  \\ 
    wasForEstimatedValue  &  Boolean &  \\ 
    isSubcontracted  &  Boolean &  \\ 
    subcontractedProportion  &  Text &  \\ 
    finalValue  &  ForeignKey & Price \\ 
    bidPrice  &  ForeignKey & Price \\ 
    wasDisqualified  &  Boolean &  \\ 
    disqualificationReason  &  Text &  \\ 
    bidder  &  ForeignKey & Bidder \\ 
\\
\multicolumn{3}{c}{\textbf{AppendicesAgreement}} \\
	\hline
	Variable & Type & Foreign table\\
	\hline
    id  &  Integer &  \\ 
    winningBid  &  ForeignKey & WinningBid \\ 
    document  &  ForeignKey & Document \\ 
\\
\multicolumn{3}{c}{\textbf{AmendmentsAgreement}} \\
	\hline
	Variable & Type & Foreign table\\
	\hline
    id  &  Integer &  \\ 
    winningBid  &  ForeignKey & WinningBid \\ 
    document  &  ForeignKey & Document \\ 
\\
\multicolumn{3}{c}{\textbf{AssignedAnnouncementIds}} \\
	\hline
	Variable & Type & Foreign table\\
	\hline
    id  &  Integer &  \\ 
    contract  &  ForeignKey & Contract \\ 
    assignedAnnouncementId  &  Text &  \\ 
\\
\multicolumn{3}{c}{\textbf{AssignedContractIds}} \\
	\hline
	Variable & Type & Foreign table\\
	\hline
    id  &  Integer &  \\ 
    contract  &  ForeignKey & Contract \\ 
    assignedContractId  &  Text &  \\ 
\\
\multicolumn{3}{c}{\textbf{Bidder}} \\
	\hline
	Variable & Type & Foreign table\\
	\hline
    id  &  Integer &  \\ 
    isConsortium  &  Boolean &  \\ 
    body  &  ForeignKey & Body \\ 
\\
\multicolumn{3}{c}{\textbf{SubcontractorsBidder}} \\
	\hline
	Variable & Type & Foreign table\\
	\hline
    id  &  Integer &  \\ 
    bidder  &  ForeignKey & Bidder \\ 
    body  &  ForeignKey & Body \\ 
\\
\multicolumn{3}{c}{\textbf{AdditionalObjedtsContract}} \\
	\hline
	Variable & Type & Foreign table\\
	\hline
    id  &  Integer & True) \\ 
    bidder  &  ForeignKey & Bidder \\ 
    document  &  ForeignKey & Document \\ 
\\
\multicolumn{3}{c}{\textbf{Bid}} \\
	\hline
	Variable & Type & Foreign table\\
	\hline
    id  &  Integer &  \\ 
    lot & ForeignKey & Lot \\ 
    bidPrice  &  ForeignKey & Price \\ 
    wasDisqualified  &  Boolean &  \\ 
    disqualificationReason  &  Text &  \\ 
    bidder  &  ForeignKey & Bidder \\ 
\\
\multicolumn{3}{c}{\textbf{BidDocuments}} \\
	\hline
	Variable & Type & Foreign table\\
	\hline
    id  &  Integer &  \\ 
    bid  &  ForeignKey & Bid \\ 
    document  &  ForeignKey & Document \\ 
\\
\multicolumn{3}{c}{\textbf{Buyer}} \\
	\hline
	Variable & Type & Foreign table\\
	\hline
    id  &  Integer &  \\ 
    mainActivity  &  Text &  \\ 
    buyerType  &  Text &  \\ 
    isPublic  &  Boolean &  \\ 
    isSubsidized  &  Boolean &  \\ 
    isSectoral  &  Boolean &  \\ 
    body  &  ForeignKey & Body \\ 
\\
\multicolumn{3}{c}{\textbf{Document}} \\
	\hline
	Variable & Type & Foreign table\\
	\hline
    id  &  Integer &  \\ 
    contract  &  ForeignKey & Contract \\ 
    title  &  Text &  \\ 
    type  &  Text &  \\ 
    url  &  Text &  \\ 
    publicationDatetime  &  Text &  \\ 
    signatureDate  &  Text &  \\ 
\\
\multicolumn{3}{c}{\textbf{OtherVersionsDocument}} \\
	\hline
	Variable & Type & Foreign table\\
	\hline
    id  &  Integer &  \\ 
    document  &  ForeignKey & Document \\ 
    otherVersion  &  ForeignKey & Document \\ 
\\
\multicolumn{3}{c}{\textbf{WinningBidPayments}} \\
	\hline
	Variable & Type & Foreign table\\
	\hline
    id  &  Integer &  \\ 
    winningBid  &  ForeignKey & WinningBid \\ 
    paymentD  &  Text &  \\ 
    amount  &  ForeignKey & Price \\ 
\\
\multicolumn{3}{c}{\textbf{RelatedAnnouncementIds}} \\
	\hline
	Variable & Type & Foreign table\\
	\hline
    id  &  Integer &  \\ 
    contractId  &  ForeignKey & Contract \\ 
    relatedAnnouncement  &  Text &  \\ 
\\
\multicolumn{3}{c}{\textbf{SubContractorsBodies}} \\
	\hline
	Variable & Type & Foreign table\\
	\hline
	id  &  Integer &  \\ 
    bidderId  &  ForeignKey & Bidder \\ 
    bodyId  &  ForeignKey & Body \\ 
\\
\multicolumn{3}{c}{\textbf{SubContractorsDocuments}} \\
	\hline
	Variable & Type & Foreign table\\
	\hline
    id  &  Integer &  \\ 
    bidderId  &  ForeignKey & Bidder \\ 
    documentId  &  ForeignKey & Document \\ 
\\
\multicolumn{3}{c}{\textbf{Corriagenda}} \\
	\hline
	Variable & Type & Foreign table\\
	\hline
    documentRef & Text &  \\ 
    publicationDate & Text &  \\ 
    correction & Text &  \\ 
\hline
\caption{Use case database description.}
\label{tab:db_schema}
\end{longtable}

\newpage

\section{Conclusions}
\label{sec:conclusion} 

In this work, we have presented a methodology to tackle the problem of web knowledge extraction. Our methodology, is an iterative machine-assisted expert-driven solution that reduces the intervention of developers. The presented solution is built around five interconnected but independent components that make possible the reuse of a large number of components in order to reduce the development time employed in any data extraction project. We describe the components of our methodology and discuss the benefits and possible drawbacks of each one.

We think our methodology differs from current solutions in the following aspects. 1) Our solution is expert-driven. This means that a domain expert can control the whole data extraction process without requiring any IT background. 2) We define a set of reusable tasks making possible a reduction in the number of working hours compared with ad-hoc implementations. 3) Our solution can be adapted to several scenarios by extending our configurable pipeline design. 4) We can extract data from several sources from structured to non-structured. 5) The expert is assisted by a document structure analysis solution that helps her in the annotation process. 6) The extracted data can be stored independently of the target database.

In order to demonstrate the benefits of our methodology, we present a real case study in the context of public procurement data. We aim to extract knowledge from several public procurement data sources from Europe into a unified model schema. This use case presents many of the common challenges to address in a web extraction project. However, we had to address many issues not mentioned in the literature such as multilingual sources, specific vocabulary not covered by any ontology, the lack of expert guidance for most of the data providers and bad or corrupted data.  During the study of this use case, we discuss the different steps carried out and we finally present an analytical description of the collected data.

\noindent \textbf{Acknowledgement.} The research is funded by H2020 DIGIWHIST project (645852). We would like to thank Christopher R\'e in Stanford University for valuable input and discussion and NLP group in University of Cambridge.

\newpage

\bibliographystyle{unsrt}
\bibliography{dwhist}

\begin{thebibliography}{10}

\bibitem{schema_org}
R.~V. Guha, Dan Brickley, and Steve Macbeth.
\newblock Schema.org: Evolution of structured data on the web.
\newblock {\em Commun. ACM}, 59(2):44--51, January 2016.

\bibitem{Raghavan:2001}
Sriram Raghavan and Hector Garcia-Molina.
\newblock Crawling the hidden web.
\newblock In {\em Proceedings of the 27th International Conference on Very
  Large Data Bases}, VLDB '01, pages 129--138, San Francisco, CA, USA, 2001.
  Morgan Kaufmann Publishers Inc.

\bibitem{memex}
\url{www.darpa.mil/program/memex}.

\bibitem{Lawrence:98}
Steve Lawrence and C.~Lee Giles.
\newblock Searching the world wide web.
\newblock {\em Science}, 280(5360):98--100, 1998.

\bibitem{Zerfos:05}
P.~Zerfos, J.~Cho, and A.~Ntoulas.
\newblock Downloading textual hidden web content through keyword queries.
\newblock In {\em Digital Libraries, 2005. JCDL '05. Proceedings of the 5th
  ACM/IEEE-CS Joint Conference on}, pages 100--109, June 2005.

\bibitem{Shestakov:2005}
Denis Shestakov, Sourav~S. Bhowmick, and Ee-Peng Lim.
\newblock Deque: Querying the deep web.
\newblock {\em Data Knowl. Eng.}, 52(3):273--311, March 2005.

\bibitem{Madhavan:2008}
Jayant Madhavan, David Ko, Lucja Kot, Vignesh Ganapathy, Alex Rasmussen, and
  Alon Halevy.
\newblock Google's deep web crawl.
\newblock {\em Proc. VLDB Endow.}, 1(2):1241--1252, August 2008.

\bibitem{scrapy}
\url{http://scrapy.org/}.

\bibitem{nutch}
\url{http://nutch.apache.org/}.

\bibitem{scrapinghub}
\url{http://scrapinghub.com/}.

\bibitem{mozenda}
\url{http://www.mozenda.com/}.

\bibitem{xquery}
\url{https://www.w3.org/XML/Query/}.

\bibitem{xslt}
\url{https://www.w3.org/TR/xslt}.

\bibitem{htmltidy}
\url{https://www.w3.org/People/Raggett/tidy/}.

\bibitem{Shen:2007}
Warren Shen, AnHai Doan, Jeffrey~F. Naughton, and Raghu Ramakrishnan.
\newblock Declarative information extraction using datalog with embedded
  extraction predicates.
\newblock In {\em Proceedings of the 33rd International Conference on Very
  Large Data Bases}, VLDB '07, pages 1033--1044. VLDB Endowment, 2007.

\bibitem{Krishnamurthy:2008}
Rajasekar Krishnamurthy, Yunyao Li, Sriram Raghavan, Frederick Reiss,
  Shivakumar Vaithyanathan, and Huaiyu Zhu.
\newblock Systemt: a system for declarative information extraction.
\newblock {\em SIGMOD}, 37(4):7--13, 2008.

\bibitem{Nakashole:2011}
Ndapandula Nakashole, Martin Theobald, and Gerhard Weikum.
\newblock Scalable knowledge harvesting with high precision and high recall.
\newblock In {\em Proceedings of the Fourth ACM International Conference on Web
  Search and Data Mining}, WSDM '11, pages 227--236, New York, NY, USA, 2011.
  ACM.

\bibitem{Dong:2014}
Xin~Luna Dong, Evgeniy Gabrilovich, Geremy Heitz, Wilko Horn, Kevin Murphy,
  Shaohua Sun, and Wei Zhang.
\newblock From data fusion to knowledge fusion.
\newblock {\em Proc. VLDB Endow.}, 7(10):881--892, June 2014.

\bibitem{Etzioni:2004}
Oren Etzioni, Michael Cafarella, Doug Downey, Stanley Kok, Ana-Maria Popescu,
  Tal Shaked, Stephen Soderland, Daniel~S. Weld, and Alexander Yates.
\newblock Web-scale information extraction in knowitall: (preliminary results).
\newblock In {\em Proceedings of the 13th International Conference on World
  Wide Web}, WWW '04, pages 100--110, New York, NY, USA, 2004. ACM.

\bibitem{Carlson:2010}
Andrew Carlson, Justin Betteridge, Bryan Kisiel, Burr Settles, Estevam
  R.~Hruschka Jr., and Tom~M. Mitchell.
\newblock Toward an architecture for never-ending language learning.
\newblock In {\em AAAI}. AAAI Press, 2010.

\bibitem{liu2004editorial}
Bing Liu and Kevin Chen-Chuan-Chang.
\newblock Editorial: special issue on web content mining.
\newblock {\em Acm Sigkdd explorations newsletter}, 6(2):1--4, 2004.

\bibitem{barathi2014structured}
B~Anantha Barathi.
\newblock Structured information extraction system from web pages.
\newblock {\em Middle-East Journal of Scientific Research}, 19(6):817--820,
  2014.

\bibitem{arasu2003extracting}
Arvind Arasu and Hector Garcia-Molina.
\newblock Extracting structured data from web pages.
\newblock In {\em Proceedings of the 2003 ACM SIGMOD international conference
  on Management of data}, pages 337--348. ACM, 2003.

\bibitem{chang2001iepad}
Chia-Hui Chang and Shao-Chen Lui.
\newblock Iepad: information extraction based on pattern discovery.
\newblock In {\em Proceedings of the 10th international conference on World
  Wide Web}, pages 681--688. ACM, 2001.

\bibitem{wang2004information}
Jiying Wang.
\newblock {\em Information extraction and integration for Web databases}.
\newblock Hong Kong University of Science and Technology (People's Republic of
  China), 2004.

\bibitem{cunningham2013getting}
Hamish Cunningham, Valentin Tablan, Angus Roberts, and Kalina Bontcheva.
\newblock Getting more out of biomedical documents with gate's full lifecycle
  open source text analytics.
\newblock {\em PLoS Comput Biol}, 9(2):e1002854, 2013.

\bibitem{Shin:2015}
Jaeho Shin, Sen Wu, Feiran Wang, Christopher De~Sa, Ce~Zhang, and Christopher
  R{\'e}.
\newblock Incremental knowledge base construction using deepdive.
\newblock {\em Proc. VLDB Endow.}, 8(11):1310--1321, July 2015.

\bibitem{crescenzi2001roadrunner}
Valter Crescenzi, Giansalvatore Mecca, Paolo Merialdo, et~al.
\newblock Roadrunner: Towards automatic data extraction from large web sites.
\newblock In {\em VLDB}, volume~1, pages 109--118, 2001.

\bibitem{selenium}
\url{http://www.seleniumhq.org/}.

\bibitem{phantomjs}
\url{http://phantomjs.org/}.

\bibitem{wordnet}
\url{https://wordnet.princeton.edu/}.

\bibitem{Navarro:2001}
Gonzalo Navarro.
\newblock A guided tour to approximate string matching.
\newblock {\em ACM Comput. Surv.}, 33(1):31--88, March 2001.

\bibitem{mongo}
\url{https://www.mongodb.org/}.

\bibitem{django_orm}
\url{https://docs.djangoproject.com/es/1.9/topics/db/}.

\end{thebibliography}



\end{document}